\documentclass[usenatbib]{mn2e}
\usepackage{psfig}
\usepackage{amssymb,amsmath}

\def\gs{\mathrel{\raise0.35ex\hbox{$\scriptstyle >$}\kern-0.6em
\lower0.40ex\hbox{{$\scriptstyle \sim$}}}}
\def\ls{\mathrel{\raise0.35ex\hbox{$\scriptstyle <$}\kern-0.6em
\lower0.40ex\hbox{{$\scriptstyle \sim$}}}}

\def\kms{\,\hbox{km}\,\hbox{s}^{-1}}
\def\Msol{\mathrel{\rm M_{\odot}}}

\def\Msolyr{\mathrel{\rm M_{\odot}yr^{-1}}}
\def\Wm2{\,\hbox{W}\,\hbox{m}^{-2}}
\def\gsim{\mathrel{\raise0.35ex\hbox{$\scriptstyle >$}\kern-0.6em\lower0.40ex\hbox{{$\scriptstyle \sim$}}}}
\def\lsim{\mathrel{\raise0.35ex\hbox{$\scriptstyle <$}\kern-0.6em\lower0.40ex\hbox{{$\scriptstyle \sim$}}}}

\begin{document}

\title[Resolved Spectroscopy of z$\sim$2--3 Lensed Galaxies]
{Resolved Spectroscopy of Gravitationally-Lensed Galaxies:
Recovering Coherent Velocity Fields in Sub-Luminous z$\sim$2--3 Galaxies}

\author[Jones et al.]
{ \parbox[h]{\textwidth}{ 
T.\,A.\ Jones$^{\, 1,*}$,
A.\,M.\ Swinbank$^{\, 2}$,
R.S. \ Ellis$^{\, 1}$,
J.\ Richard$^{\, 2}$,
D.\,P.\ Stark$^{\, 3}$
}
\vspace*{6pt} \\
$^1$Astronomy Department, California Institute of Technology, 249-17, Pasadena, CA 91125, USA \\
$^2$Institute for Computational Cosmology, Department of Physics, Durham University, 
South Road, Durham, DH1 3LE, UK \\
$^3$Kavli institute for Cosmology and Institute of Astronomy, University of Cambridge, Madingley Road, Cambridge, CB3 0HA, UK \\
$^*$Email: tajones@astro.caltech.edu \\
}

\maketitle 

\begin{abstract}
We present spatially-resolved dynamics for six strongly lensed star-forming galaxies at $z = 1.7$--3.1, each enlarged by a linear magnification factor $\sim\times$8.  Using the Keck laser guide star AO system and the OSIRIS integral field unit spectrograph we resolve kinematic and morphological detail in our sample with an unprecedented fidelity, in some cases achieving spatial resolutions of $\simeq$100\,pc.  With one exception our sources have diameters ranging from 1--7 kpc, integrated star formation rates of $2-40\,\Msolyr$ (uncorrected for extinction) and dynamical masses of $10^{9.7-10.3} \Msol$. With this exquisite resolution we find that four of the six galaxies display coherent velocity fields consistent with a simple rotating disk model. Our model fits imply ratios for the systemic to random motion, $V_c \sin{i} / \sigma$, ranging from 0.5--1.3 and Toomre disk parameters $Q < 1$.  The large fraction of well-ordered velocity fields in our sample is consistent with data analyzed for larger, more luminous sources at this redshift. We demonstrate that the apparent contradiction with earlier dynamical results published for unlensed compact sources arises from the considerably improved spatial resolution and sampling uniquely provided by the combination of adaptive optics and strong gravitational lensing.  Our high resolution data further reveal that all six galaxies contain multiple giant star-forming H{\sc ii} regions whose resolved diameters are in the range 300 pc -- 1.0 kpc, consistent with the Jeans length expected in the case of dispersion support. From the kinematic data we calculate that these regions have dynamical masses of $10^{8.8-9.5} \Msol$, also in agreement with local data. However, the density of star formation in these regions is $\sim$100$\times$ higher than observed in local spirals; such high values are only seen in the most luminous local starbursts. The global dynamics and demographics of star formation in these H{\sc ii} regions suggest that vigorous star formation is primarily governed by gravitational instability in primitive rotating disks. The physical insight provided by the combination of adaptive optics and gravitational lensing suggests it will be highly valuable to locate many more strongly-lensed distant galaxies with high star formation rates before the era of the next-generation ground-based telescopes when such observations will become routine.

\end{abstract}


\section{Introduction}
\label{secintro}
Studies of star-forming galaxies at high redshift ($z > 1.5$) over the
past decade have mapped the demographics of the populations as a whole
yielding valuable data on their integrated star formation rates,
stellar and dynamical masses, metallicities and morphologies
\citep{Shapley01,Shapley03,Reddy04,Erb06, Law07a}. From these data, a consistent
picture of the development of the comoving density of star formation
and its contribution to the present day stellar density has emerged
\citep{Dickinson03, Hopkins06}. Following a rapid rise in activity
at early times, corresponding to the redshift range $2<z<6$, the star formation
rate has decline markedly over the past 8 Gyr corresponding to the interval
$0<z<1$. Of particular interest is the
relationship between sources observed at the time close to the peak
epoch of star formation, $z = 2-3$ and the population of massive
galaxies observed today.

Recent theoretical work has focused on the mechanisms via which early
galaxies assemble their stars and evolve morphologically to the
present day Hubble sequence. Minor mergers and cold stream gas
accretion have emerged as a possible means of building disks,
central bulges and elliptical galaxies from early star forming
Lyman break galaxies \citep{Dekel09, Brooks09}.
These models of galaxy evolution rely on simple descriptions of
complex physical processes such as gas cooling, star formation and
feedback mechanisms.  It is therefore crucial to undertake relevant
observations to constrain these processes.  The most direct
observational route to making progress is high quality resolved
spectro-imaging of early galaxies which can be used to determine both
their dynamical state and the distribution of star-formation.

Integral field unit (IFU) spectroscopy of nebular emission lines is
now yielding valuable kinematic data for a growing sample of luminous
high-redshift star forming galaxies \citep{Genzel06, Genzel08,
Forster06, Forster09, Law07, Law09}. Although the galaxies observed
so far do not yet in any way constitute a complete well-defined
sample, some important results have emerged. The data reveal a mix of
dispersion-dominated systems, rotating systems, and major mergers with
the common observation that all galaxies studied so far show
relatively high velocity dispersions of $\sim 50-100$ km\,s$^{-1}$. A
larger fraction of the more massive ($M_{dyn} \gtrsim 10^{11}
M_{\odot}$) galaxies appear to be rotationally supported, with lower
mass galaxies ($M_{dyn} \lsim 10^{10}M_{\odot}$) tending to have 
dispersion-dominated kinematics \citep{Law09}.

However, these studies are hampered by the poor spatial resolution
inherent in studies of distant star-forming galaxies whose typical
sizes are $\simeq1$--5\,kpc. Even with adaptive optics on a 8--10 meter
aperture, the physical resolution is limited to FWHM $\gtrsim 1$ kpc
which means only a few independent resolution elements. Consequently
it is unclear whether the velocity shear observed in the smaller
2--5\,kpc sources arises from rotation or merging. Moreover, although
there are suggestions that distant star-forming regions have sizes
much larger than those of local H{\sc ii} regions \citep{Elmegreen05},
the claim remains uncertain since the distant regions are not properly
resolved.  Thus, the diffraction limit of the current generation of
optical/infrared telescopes limits progress, even with all the
adaptive optics tools of the trade.

Fortunately, sources which undergo strong gravitational lensing are
enlarged as seen by the observer and can thus be studied at much
higher physical resolution in their source plane \citep{Nesvadba06,
  Swinbank07}. This improvement in physical resolution enables us to
better distinguish various forms of velocity field and to examine the
properties of star-forming giant H{\sc ii} regions.  A convincing
demonstration of the benefits of lensing as applied to this topic was
the work of \cite{Stark08} which analyzed IFU data on a $z=3.07$
galaxy magnified in area by a factor of 28$\times$ demonstrating
resolved dynamics on $\simeq$100 pc scales. This represents an order
of magnitude improved sampling compared to the earlier studies cited
above.

Here we present IFU observations of a further five strongly lensed
galaxies which we add to the source studied by \cite{Stark08}. In
addition to studying the well-sampled kinematic structure of a
population of sub-L$^{\ast}$ star-forming galaxies at $z\simeq2$, our
fine resolution allows us to examine the size, dynamical mass, and
luminosity of typical star-forming regions, thereby probing the
conditions in which most of the star formation takes place. We find
that all of our galaxies contain multiple, distinct star forming
regions.  These results offer a preview of the science which will be
possible with 30 meter class optical/near-IR telescopes with a
diffraction limit comparable to the resolution of our data; they
provide the most detailed view to date of typical high-redshift star
forming galaxies.

A plan of the paper follows. In $\S$2 we discuss our selection of
targets and Keck observations, including the production and use of
gravitational lens modeling essential for achieving a high resolution
in the source plane. In $\S$3 we analyze our spectroscopic results,
discussing each object in turn and summarizing the kinematic data. In
$\S$4 we examine the physical scale of our star forming regions in
comparison to those seen locally. We summarize our conclusions in
$\S$5.  Throughout we use a {\it WMAP} cosmology \citep{Spergel03}
with $\Omega_{\Lambda}$=0.73, $\Omega_{m}$=0.27, and
H$_{0}$=72\,km\,s$^{-1}$\,Mpc$^{-1}$.  All quoted magnitudes are in
the AB system unless otherwise noted.

\section{Observations and Data Reduction}
      
\subsection{Target Selection}

Ideally, in considering the selection of high redshift galaxies for
more detailed resolved studies, care would be taken to construct a
mass-limited sample, perhaps defined additionally according to the
integrated or specific star formation rate. In practice, even with 10
meter class telescopes, studies are limited by the surface brightness
distribution of gaseous emission lines as discussed by earlier workers
\citep{Genzel06, Genzel08, Forster06, Forster09, Law07, Law09}. In
examining {\em gravitationally-lensed} systems, which have the unique
advantages of probing to less luminous (and presumably more typical)
systems, and offering an increased spatial resolution in the source
plane, an additional criterion is the reliability of the mass model
used to invert the observed data into that of the source plane.

Over the past decade, using the growing archive of {\it HST} ACS and
WFPC-2 images of rich clusters (e.g. \citealt{Santos04, Sand05,
  Stark07}) and through specific projects such as the MAssive Cluster
Survey 
(MACS, \citealt{Ebeling01}) and Local Cluster Substructure Survey (LoCuSS, \citealt{Smith05}),
we have
studied a large sample of lensing clusters and undertaken a search
within these {\it HST} images for promising magnified galaxies (arcs)
at high redshift. The clusters presented herein do not in any way represent a
well-defined sample and, inevitably, there are biases to those which
act as spectacular gravitational lenses.

The distant sources described here were located and assessed via two
sequential programs. Firstly, a comprehensive multi-object
spectroscopic campaign involving Keck and the VLT was used to verify
the redshift of the candidate arcs as well as to determine the
redshifts of other multiply-imaged sources in each cluster in order to
construct robust mass models (e.g. \citealt{Richard07}). As the
follow-up spectroscopy was undertaken at optical wavelengths, for
those sources beyond $z\simeq$2, this provided valuable insight into
the strength and distribution of Ly$\alpha$ emission. However, as with
previous work (e.g. \citealt{Shapley03}) we found that the Ly$\alpha$
line did not necessarily give a good indication of the relative
strengths and extended nature of those nebular emission lines ([O{\sc ii}],
H$\beta$, [O{\sc iii}], H$\alpha$) critical for resolved studies with laser
guide-star assisted adaptive optics (LGSAO).  Accordingly, we used the
Keck II spectrograph, NIRSPEC, in seeing limited mode in a second
campaign to `screen' all promising targets in advance of undertaking
the more demanding resolved studies with the LGSAO-fed integral field
unit spectrograph OSIRIS. Once the redshift of the lensed source was
known from optical spectroscopy, we usually followed up the source in
more than one near-infrared line, in order to ascertain which would be
the most economic tracer of the velocity field.

\begin{figure*}
\centerline{\psfig{file=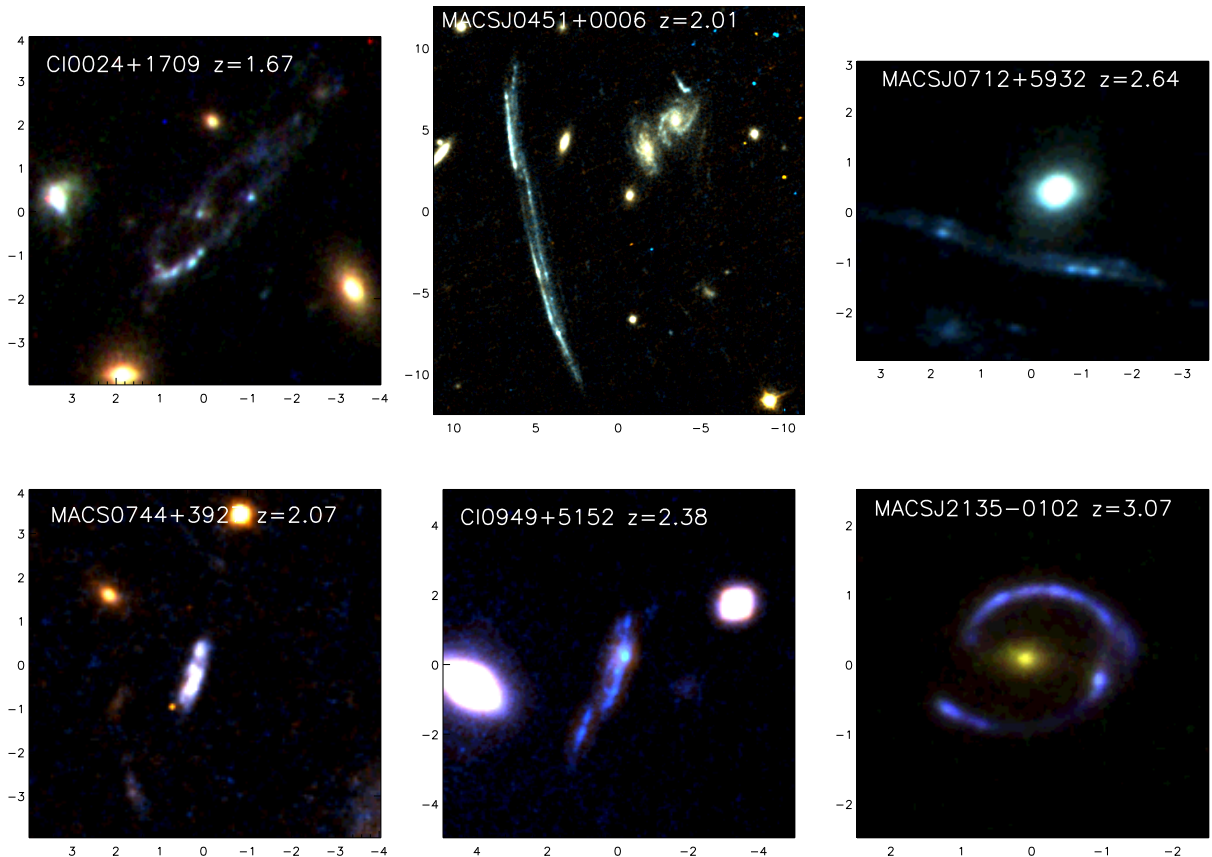,angle=90,width=6.5in}}
\caption{False color ACS images of the sample derived from the multi-band data
referred to in Table 1. The xy scale is in arcseconds.}
\label{fig:hst_montage}
\end{figure*}

About 75\% of our initial sample of 30 $z>1.5$ arcs have so far been
screened and have suitably close reference stars for tip-tilt
correction. In order to achieve an adequate signal to noise with
OSIRIS, we used the NIRSPEC data to limit the sample to those with an
emission line surface brightness $\gtrsim 10^{-16}$
erg\,s$^{-1}$cm$^{-2}$arcsec$^{-2}$ in a region uncontaminated by
night sky emission. As a guide, this surface brightness criterion
corresponds to a SFR density $> 0.3\,M_{\odot}$yr$^{-1}$kpc$^{-2}$ for
H$\alpha$ at $z=2$.  Our sample of 6 sources is drawn from a list of
$\simeq20$ objects fulfilling these criteria.  A full account of the
NIRSPEC screening program and a discussion of the distribution of line
strengths found in that survey is given by \cite{Richard09}. Key
properties for the sample of 6 objects, which includes the source
discussed by \cite{Stark08}, are given in Table 1.  The ACS images of
our lensed targets are given in Figure~\ref{fig:hst_montage} and we
briefly discuss these below.

{\it Cl0024+1654} Deep optical and near-infrared imaging of the
$z=0.39$ cluster Cl\,0024+1654 with {\it HST} reveals five distinct
images of a background galaxy originally identified by its blue color
and clumpy ring-like morphology (\citealt{Colley96},
Figure~\ref{fig:hst_montage}).  Optical spectroscopy established a
redshift of $z=1.675$ from interstellar absorption lines
\citep{Broadhurst00}.  Near-infrared spectroscopy shows bright,
spatially extended H$\alpha$ emission at $z=1.6795$ (Richard et
al. 2009).  To map the velocity field, we observed the counter image 
northwest of the cluster with OSIRIS around the redshifted H$\alpha$ and
[N{\sc ii}]$\lambda$6583 emission.

{\it MACS0451+0006} The {\it HST} ACS $V$- and $I$-band imaging of
MACSJ0451+0006 shows a striking, multiply-imaged arc $\simeq$30
arcseconds east of the brightest cluster galaxy
(Figure~\ref{fig:hst_montage}). Follow-up spectroscopy with the FORS
multi-object spectrograph on the VLT confirmed a redshift of $z=2.008$
via the identification of Ly$\alpha$ and UV ISM absorption lines
(Si{\sc ii}$\lambda$1260, O{\sc i}$\lambda$1303, C{\sc iv}$\lambda$1549 and
Si{\sc iv}$\lambda$1400) and He{\sc ii}$\lambda$1640 and C{\sc
  iii}]$\lambda$1908.7 in emission (PID: 078.A-0420).  Follow-up
observations targeting the nebular H$\alpha$ emission with the NIRSPEC
longslit spectrograph on Keck II showed strong, spatially extended
H$\alpha$ at a redshift of 2.0139$\pm$0.0001 \citep{Richard09}.

{\it MACS\,J0712+5932} {\it HST} $V$- and $I$-band imaging of
MACS\,J0712+5932 reveals a prominent triply-imaged background
galaxy.  Each of the galaxy images clearly comprises two
prominent clumps with at least one fainter region with surface
brightness $\mu_V>23$\,mags\,arcsec$^{-2}$ within a diffuse halo
(Figures~\ref{fig:hst_montage},\ref{fig:osiris_montage}).  Follow-up
near-infrared spectroscopy \citep{Richard09} shows strong nebular
H$\alpha$, H$\beta$, and [O{\sc iii}]$\lambda\lambda5007,4959$
emission at $z = 2.6462$, concentrated in the UV-bright regions.

{\it MACS\,J0744+3927} {\it HST} $V$- and $I$-band imaging of
MACS\,J0744+3927 reveals an arc approximately 15$"$ west of the cluster
center. The optical redshift $z=2.207$ obtained via identification of
strong Ly$\alpha$ absorption and multiple rest-frame UV absorption
lines was later confirmed by the detection of strong H$\alpha$ and
[N{\sc ii}] emission at $z=2.209$ with NIRSPEC.

{\it Cl0949+5153} The {\it HST} $V$-band image of Cl\,0949+5153 shows
a clumpy, elongated arc resolved into two diffuse components
(Figure~\ref{fig:hst_montage}).  Further
spectroscopy around the redshifted nebular emission lines shows
extended H$\alpha$, H$\beta$ and [O{\sc iii}]$\lambda\lambda5007,4959$
emission at $z=2.393$.

{\it MACS\,J2135-0102} For completeness, we also include in our sample
discussion of the velocity field of the lensed Lyman Break Galaxy
(LBG) at $z=3.07$ in MACS\,J2135-0102 (also known as the ``Cosmic
Eye'' due to its lensed morphology; \citealt{Stark08}).  Multi-wavelength
studies have shown that the source is a L$^{*}$ Lyman-break galaxy
with a stellar age of 80--300\,Myr, star formation rate of
SFR\,=\,40--60\,$\Msolyr$, and stellar mass of
M$_{\star}=6\pm2\times10^9 \Msol$ \citep{Coppin07, Siana09}.

\subsection{OSIRIS Observations and Data Reduction}

Detailed 2-D spectroscopic observations of the targets listed in Table
1 were made with the near-infrared integral field spectrograph OSIRIS
\citep{Larkin06} on the 10 meter Keck II telescope using the laser
guide star adaptive optics (LGSAO) system \citep{Wizinowich06} to
correct for atmospheric distortion. A suitably bright star ($R<17$)
within $\sim 50''$ of the target was used for tip-tilt
correction. We used the 100\,milli-arcsecond pixel scale in all
observations which provides a field of view of at least
3.2$"\times$6.4$"$. Narrow band $H$ and $K$ filters (Table~1) were selected to
target the H$\alpha$ or [O{\sc iii}]$\lambda 5007$ emission line at a
spectral resolution $R \simeq 3600$, corresponding to FWHM $\simeq 6$
\AA\ in the K band. MACS\,J2135-0102 and MACS\,J0451+0006 were observed
on 2007 September 02 and 03 whilst the remaining targets were observed
between 2008 November 27 -- 30. %For MACS\,J0451+0006, the southern
%region of the arc was observed in September 2007 with LGSAO and the
%northern region was observed in November 2008 with no adaptive optics
%correction. In this paper we present only the September 2007 data as
%it has superior resolution and covers a greater fraction of the galaxy
%source plane, including the entire region probed by the non-AO data.

All observations were taken in $0.6-1.3"$ seeing. Before observing
each arc, we took short exposures of the tip-tilt reference star to
center the IFU pointing. The tip-tilt exposures were also used for
flux calibration and to calculate the point spread function. Gaussian
fits to the point spread functions of the tip-tilt stars yield
0.13--0.20$"$ FWHM resolution for the 100 milliarcsecond pixel scale.
With smaller pixel scales, the resolution delivered by the LGSAO
system was 0.11$"$ FWHM. Observations of each target were done in a
standard ABBA position sequence to achieve good sky subtraction. In
the case of Cl0024+1709 we chopped 8$"$ to sky, whilst in the
remaining cases we chopped the galaxy within the IFU. Individual
exposures were 600--900 seconds and each observing block was
2.4--3.6\,ks which was typically repeated three to six times. The
total integration time for each object is given in Table 1.

Our data reduction methods closely followed those described in detail
by \cite{Stark08}. We used the OSIRIS data reduction pipeline
\citep{Larkin06} to perform sky subtraction, spectral extraction,
wavelength calibration, and form the data cube. To accurately combine
the individual data cubes, we created images of the integrated emission
line and used the peak intensity to centroid the object. We then
spatially aligned and co-added the individual data cubes to create the
final mosaic. Flux calibration was performed by equating the flux
density of the tip-tilt stars measured from 2MASS photometry with the
observed OSIRIS spectra. We estimate that the uncertainty in
flux calibration is typically 10\%.
%, although fluctuations in the LGSAO
%point-spread function may introduce additional systematic
%uncertainty of up to $\sim 30$\% (see discussion in \citealt{Law07}).

\begin{table*}
\begin{center}
{\footnotesize
{\centerline{\sc Table 1.}}
{\centerline{\sc Target List}}
\begin{tabular}{lccccccccc}
\hline
\noalign{\smallskip}
Name             & $\alpha_{J2000}$ & $\delta_{J2000}$ & $z$  & t$_{exp}$ & Emission lines & $\mu_1 \times \mu_2$ & $\mu$ & FWHM & {\it HST} ACS \\
                 & $^{h\, m\, s}$   & $^{\circ}$ $^{'}$ $^{''}$ & & (ks) & and OSIRIS filter & & & (pc) & photometry \\
\hline
Cl\,0024+1709    & 00\,26\,34.43 & +17\,09\,55.4 & 1.680 & 16.5 & H$\alpha$, [N{\sc ii}]; Hn5 & 0.8 $\times$ 1.7 & $1.38 \pm .15$ & 820 & B,V,r,i,z \\
MACS\,J0451+0006 & 04\,51\,57.27 & +00\,06\,20.7 & 2.014 & 14.4 & H$\alpha$, [N{\sc ii}]; Kn1 & 1.3 $\times$ 37.0 & $49 \pm 11$ & 60 & V,I \\
MACS\,J0712+5932 & 07\,12\,17.51 & +59\,32\,16.3 & 2.648 & 16.2 & H$\alpha$, [N{\sc ii}]; Kc5 & 1.5 $\times$ 18.7 & $28 \pm 8$ & 90 & V,I \\
MACS\,J0744+3927 & 07\,44\,47.82 & +39\,27\,25.7 & 2.209 & 14.4 & H$\alpha$, [N{\sc ii}]; Kn2 & 1.9 $\times$ 8.6& $16 \pm 3$ & 310 & V,I \\
Cl0949+5153      & 09\,52\,49.78 & +51\,52\,43.7 & 2.394 & 19.2 & [O{\sc iii}]; Hn4 & 1.2 $\times$ 6.0 & $7.3 \pm 2.0$ & 350 & V \\
MACS\,J2135-0102 & 21\,35\,12.73 & -01\,01\,43.0 & 3.074 & 21.6 & H$\beta$, [O{\sc iii}]; Kn1 & 3.5 $\times$ 8.0 & $28 \pm 3$ & 120 & V,I \\
\hline
\label{table:obslog}
\end{tabular}
}
\caption{Notes: Gravitational lens modeling is discussed in Section
  2.3. The lensing amplification is non-uniform and highly
  directional, thus we give the typical linear magnifications $\mu_1$,
  $\mu_2$ along the minor and major lensing axes as well as the
  overall flux magnification $\mu$. The source plane resolution refers
  to the typical FWHM of a point source in the direction of highest
  magnification. The photometric bands B, V, r, i, I, z are ACS
  filters F435W, F555W, F625W, F775W, F814W, and F850W respectively.}
\end{center}
\end{table*}

\subsection{Gravitational Lens Modeling}

In order to investigate the source plane properties we must first
correct for the distortion and magnification by the cluster lens. We
summarize here the ingredients necessary to construct the relevant
cluster mass models. We will follow the methodology defined by earlier
relevant articles \citep{Kneib93, Kneib96, Smith05, Jullo07} within
which further details can be found.
 
Our basic approach is to use the code {\tt Lenstool}\footnote{\tt
  http://www.oamp.fr/cosmology/lenstool} \citep{Kneib93, Jullo07} to
constrain a parameterized model of the dark matter distribution. For
each cluster, the model comprises two components: 1 or 2 cluster-scale
dark matter halos, parametrized with a dual pseudo-isothermal
elliptical mass distribution (dPIE, \citealt{Eliasdottir07}), and
$\sim$50 galaxy-scale dPIE dark matter halos, centered on massive
cluster members occupying the strong lensing region in order to
account for the presence of substructure. These galaxy-scale halos are
assumed to have mass properties that follow a scaling relation based
on the luminosity of the underlying galaxy, assuming a constant
mass-to-light ratio (e.g. \citealt{Smith05}).
 
Details on the construction of each gravitational lens model are given
in \cite{Dye07} for MACS\,J2135-0102, \cite{Limousin09} for Cl0024+1709
and MACS J0744+3927 and \cite{Richard09} for the remaining 3 sources
in Table 1. Strong lensing constraints originate from the
identification of 2-5 multiply-imaged systems per cluster within the 
various ACS images. We use the astrometric positions and spectroscopic/photometric
redshifts of these sources\footnote{Spectroscopic redshifts are
  available for a majority of the systems} as individual constraints
to derive best fit parameters on the mass distribution. {\tt Lenstool}
uses a Markov Chain Monte Carlo (MCMC) sampler to derive a family of
mass models suitably fitting the strong lensing constraints, and we
use these to derive the uncertainty on each parameter of the mass
distribution.

For each of the sources presented in this paper, the best model was
then used to derive the geometrical transformation necessary for
mapping the source plane coordinates into the image plane. This
transformation enables us to reconstruct the {\it HST} morphology and
H$\alpha$ emission line images in the source plane assuming
conservation of surface brightness. The spot magnification $\mu_{xy}$
and its associated error are computed with {\tt Lenstool} at different
positions across the object, using the family of mass models from the
MCMC sampler. We can verify this value by computing the total
magnification from the ratio of the sizes (or equivalently, the total
fluxes) between the image and its source plane reconstruction. As the
magnification factor is not isotropic, the angular size of each image
is more highly stretched along a specific orientation
(Figure~\ref{fig:hst_montage}) thus affecting our source plane
resolution. The linear factors $\mu_1$ and $\mu_2$ of the
magnification (with $\mu=\mu_1\times\mu_2$) together with their
associated errors are listed in Table 1. A detailed illustration of
the uncertainties of the mass modeling method for MACS\,J2135-0102 is
given in \cite{Stark08}.

A key parameter in our analysis is the {\it physical resolution} we
achieve in the source plane for each target. To measure this, we use
observations of the tip-tilt reference stars. These serve this purpose
well as they are point sources observed with conditions and an
instrumental configuration identical to those of our distant
targets. We ``reconstruct'' these stars in the source plane
as if they were located at the arc position using the
same transformation as for the lensed galaxies, and fit a bivariate
Gaussian to the point spread function in order to determine the source
plane resolution. The typical FWHM of each arc in the direction of
highest magnification is listed in Table 1 and, with the
exception of Cl0024+1709 which is not highly-magnified, varies from
60--350\,pc with a mean of 200\,pc.

\section{Analysis}

First, we reconstruct the reduced, flux-calibrated data cubes to the
source plane using transformations from the gravitational lens models
described in \S2.3. The resulting data cubes were binned such that
each spatial pixel corresponded to 0.5--1 FWHM resolution elements
along the direction of highest magnification (Table~1). 
We fit Gaussian profiles to the
strongest emission line (H$\alpha$ or [O{\sc iii}]) at each spatial
pixel using a weighted $\chi^2$ minimization procedure and determine
the two-dimensional intensity, velocity, and velocity dispersion maps.
To compute the emission line fits we first subtracted the median value
at each spatial pixel to remove any source continuum and residual sky
background.  A blank region of sky within each data cube was used to
determine the sky variance spectrum $V(\lambda)$. The spectra to be
fit are weighted by $w(\lambda) = V^{-1}$ appropriate for Gaussian
noise, so that regions of higher noise (e.g. strong sky emission
lines) do not cause spurious fits. We compute the $\chi^2$ statistic
for the best-fit Gaussian as well as a for a featureless spectrum
($f(\lambda) = 0$) and require a minimum improvement over the fit with
no line of $\Delta \chi^2 = 16$--25 for the various arcs 
(i.e. 4--5$\sigma$ emission line
detection).  If this criterion was not met, we averaged the surrounding
3$\times$3 spatial pixels to achieve higher signal-to-noise.  No fit
was made if the 3$\times$3 averaging still failed to produce the
minimum $\Delta \chi^2$ improvement.  We calculated the formal
1$\sigma$ error bounds by perturbing the Gaussian fit parameters until
the $\chi^2$ increases by 1 from the best-fit value.
%These errors are accurate since the weighting results
%in best-fit reduced $\chi^2_{\nu}$ values close to unity, with mean
%$\chi^2_{\nu}$ ranging from 0.6--1.3 for the various arcs. We computed
%The velocity, dispersion, and intensity of nebular emission from the
%best-fit Gaussian position, width, and amplitude respectively. 
In all following sections, we have deconvolved the line widths with
the instrumental resolution ($R\simeq3600$) by subtracting the
instrumental resolution in quadrature from the best-fit Gaussian
$\sigma$.  The resulting source plane intensity, velocity, and
dispersion fields for the entire sample are shown in
Figure~\ref{fig:osiris_montage} and demonstrate detailed kinematic and
morphological properties on scales down to 100--200 pc.

%Emission line fits to the triply-imaged arcs MACS\,J0451+0006 and
%MACS\,J0712+5932 were done in the image plane to provide the highest
%possible resolution. Reconstructing the entire triple-imaged systems
%increases the signal-to-noise by a factor $\sim$\,1.5 at the cost of
%decreased source plane resolution due to spatially varying
%magnification across the arcs. In both sources the multiply imaged
%regions show consistent velocity fields at varying resolution, as
%expected. We therefore mapped the best-fit emission line parameters
%from only the most highly-magnified single image of each target into
%the source plane.  

From the source plane nebular emission line intensity and dynamics we
estimate the size, dispersion and dynamical mass of each galaxy. The
size is calculated as the maximum diameter from pixels with successful
emission line fits, roughly equivalent to a  major axis
diameter with limiting isophote $\sim10^{-16}$\,erg\,s$^{-1}$\,cm$^{-2}$\,arcsec$^{-2}$. The uncertainty is dominated by errors in the lensing
magnification and is $\leq20$\%.  We estimate the
global average velocity dispersion of each galaxy as the flux-weighted
mean of the fit pixels, $\sigma_{mean}=\sum\sigma_{pix}I_{pix}$ with
typical uncertainty of $\sim5$\% estimated from errors in the fit
parameters and the spread of $\sigma$ in individual pixels.  This
measurement of the dispersion is not affected by resolved velocity
gradients. The six galaxies in our sample all have large
$\sigma_{mean}=50-100\,\kms$, which is consistent with other resolved
observations of non-lensed galaxies at $z\gsim2$
(e.g. \citealt{Forster09, Law09, Lehnert09})

\subsection{Kinematic Modeling}

To test whether the kinematics of each galaxy are consistent with a rotating system, and to estimate the inclination of any disks, we construct and fit simple disk models to the observed velocity fields. The disk model for MACS\,J2135-0102 is discussed in detail by \cite{Stark08}, and we follow a similar method for the rest of the sample. We use an arctangent function to estimate the circular velocity as a function of radius,
\begin{equation}
V(R) = V_0 + \frac{2}{\pi} V_c \arctan{\frac{R}{R_t}}
\label{eq:rotcurve}
\end{equation}
which \cite{Courteau97} showed to be an adequate simple fit to galaxy
rotation curves in the local universe.  The disk models contain seven
parameters: inclination $i$, position angle $\theta$, coordinates
($\alpha$, $\delta$) of the disk center, scale radius $R_t$,
asymptotic velocity $V_c$, and systemic velocity $V_0$.

To test how well these simple models can describe the data, we
constructed velocity fields covering a large range of parameter space
for each galaxy.  From the disk center, inclination, and position
angle we compute the disk radius at each pixel in the source-plane
velocity maps.  We then calculate the circular velocity from
Equation~\ref{eq:rotcurve}, and correct for the azimuthal angle and
inclination to obtain the observed velocity field of the model disk.
To simulate lensing distortion, the models were then convolved with an
elliptical point spread function estimated as the best 2D Gaussian fit
to the reconstructed tip-tilt reference star images.  To estimate the
goodness-of-fit we compute the $\chi^2$ statistic using the 1$\sigma$
error from emission line fits. 
The emission line fits routinely yield
a reduced $\chi_{\nu}^2$ close to unity indicating that these errors
are a good estimate of the true uncertainty.
To estimate uncertainty in the parameters we perturb the model until
the $\chi^2$ increases by one standard deviation from the best fit.
Best-fit $\chi_{\nu}^2$ values for each galaxy are given in Table~2,
except for Cl\,0949+5152 for which no reasonable fit was found.
Velocity contours for the best-fit disk models are shown in Figure~\ref{fig:osiris_montage}.

For galaxies which can be reasonably well described with disk-like
kinematic structure, we extract the rotation velocity, inclination and
disk position angle (major axis) from the model.  The latter two
values are usually estimated from morphology, but such methods are not
necessarily accurate in this case due to the lensing distortion and
asymmetric light distribution of these objects.  We therefore adopt
the inclination and position angle from best-fit disk models.  In
Figure~\ref{fig:rotcurves} we show the one-dimensional velocity
and dispersion profiles extracted from the data along a slit aligned
with the disk position angle (or with the morphological major axis for
the merger Cl\,0949+5153).  This allows us to better quantify the
kinematic structure and, particularly, to estimate an approximate
rotational speed $V_{max}$ for each galaxy. The circular velocity is a
parameter in the disk models, however, in some objects the velocity
field does not reach an asymptotic value (e.g. MACS\,J0744+3927).  In
such cases the best fit value of $V_c$ does not necessarily accurately
represent the observed range of velocity.  However, 4/5 galaxies
in addition to MACS\,J2135-0102
display a coherent and roughly monotonic velocity curve consistent
with the disk model.  From these we estimate $V_{max}\sin{i}$ as half
of the range in the one-dimensional model profile in the region
detected in H$\alpha$ emission, which provides a more reliable measure
of the observed velocity range than the best-fit $V_c\sin{i}$.  Values
of $V_{max}\sin{i}$ as well as the best-fit inclination from disk
models are given in Table~2.

To estimate the dynamical mass for each galaxy, we adopt two
approaches.  First, we use the velocity dispersion and maximum extent
of the H$\alpha$ emission and calculate $M_{dyn}(R)=C\,R\,\sigma_{mean}^2/G$
with $C = 5$ appropriate for a sphere of uniform density and the
radius is taken as half the maximum diameter.  The
geometrical factor $C$ is likely between 3 and 10 for these highly
turbulent galaxies \citep{Erb06}.  The dynamical mass is
likely uncertain to within a factor 2$\times$ due to the assumed
geometry and exclusion of rotationally-supported mass, whereas random 
uncertainty propagated from $R$ and $\sigma_{mean}$ is $<25$\%.  
We also use the velocity gradient to estimate the dynamical mass
via $M_{dyn} \sin^2{i}=R\,(V_{max}\sin{i})^2/G$ but
note that the maximum velocity shear in the sample gives
$\Delta V/2\sigma_{mean}\lsim1.5$ for all galaxies, so the rotational
mass is less than half of the dispersion-supported mass 
before correcting for inclination.  In future sections,
we therefore typically adopt the dynamical mass calculated from
velocity dispersion, noting that the total uncertainty is roughly a
factor of two dominated by the assumed $C = 5$ with an additional
systematic error due to the ignored rotational support.  Adopted
values of the diameter, dispersion, and dynamical mass are given in
Table~2.

\begin{figure*}
\centerline{\psfig{file=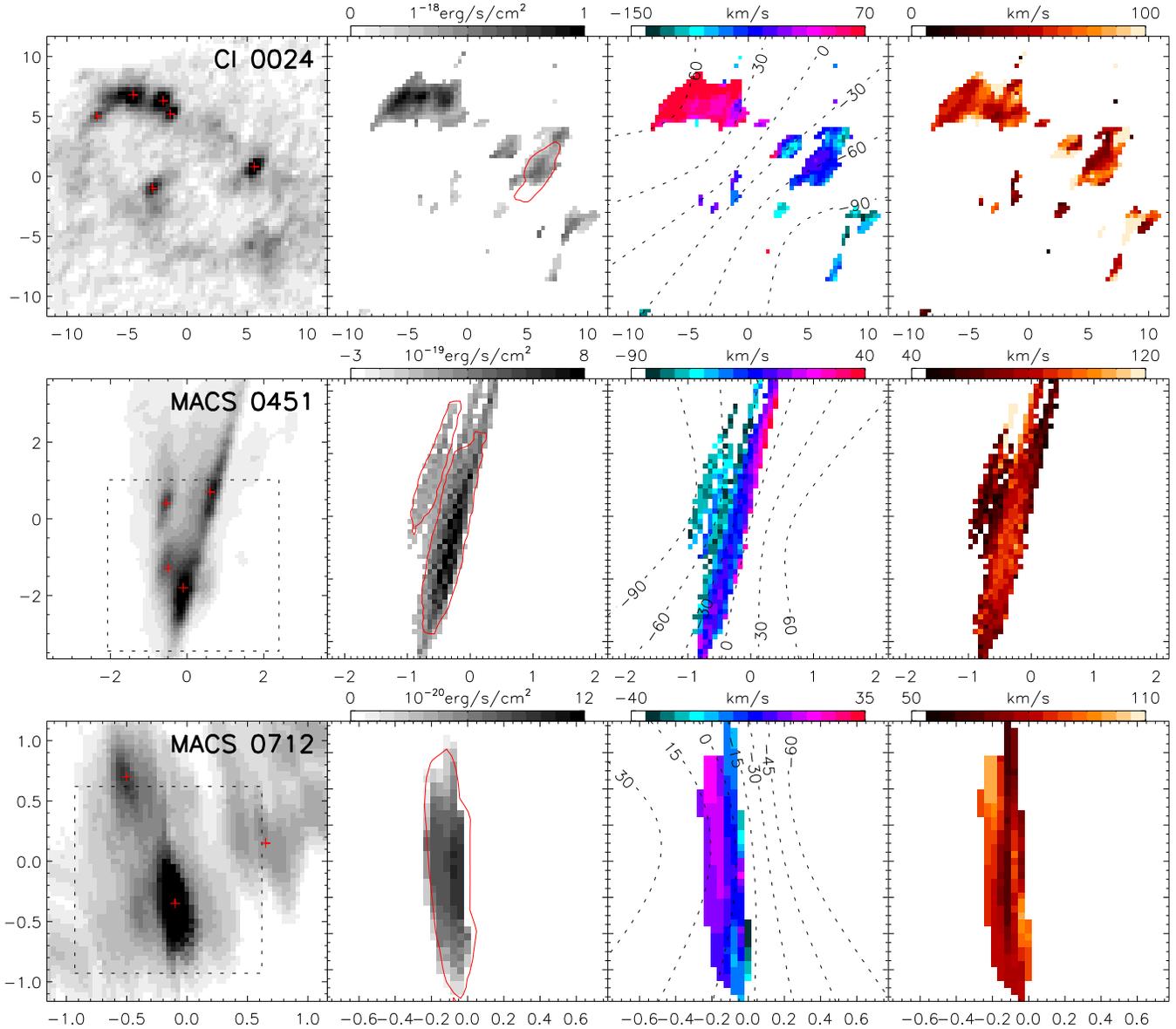,angle=0,width=7in}}
\caption{Source plane properties of the lensed galaxy sample. From
  left to right in each row: ACS broad-band emission, line intensity,
  velocity and velocity dispersion derived from the OSIRIS nebular
  emission line data. Axes are in kpc with 1--2 pixels per OSIRIS
  resolution element. Each galaxy shows morphological and kinematic
  structure on scales down to $\sim 100-200$ parsecs. The nebular line
  flux distribution is similar to the rest-UV morphology with multiple
  resolved clumps, shown as red crosses on the ACS images. 
  Red contours on the line intensity maps denote clumps
  which are spatially extended and unconfused at the $\simeq3\sigma$
  flux isophote in narrow-band emission line images, as described in
  \S4. Best-fit disk model contours are shown on the velocity maps for those galaxies 
  whose dynamics are consistent with rotation (\S3.1). Dashed boxes in the ACS images of MACS\,J0451 and MACS\,J0712 indicate the smaller regions for which
OSIRIS data are displayed.}
\label{fig:osiris_montage}
\end{figure*}

\addtocounter{figure}{-1}
\begin{figure*}
\centerline{\psfig{file=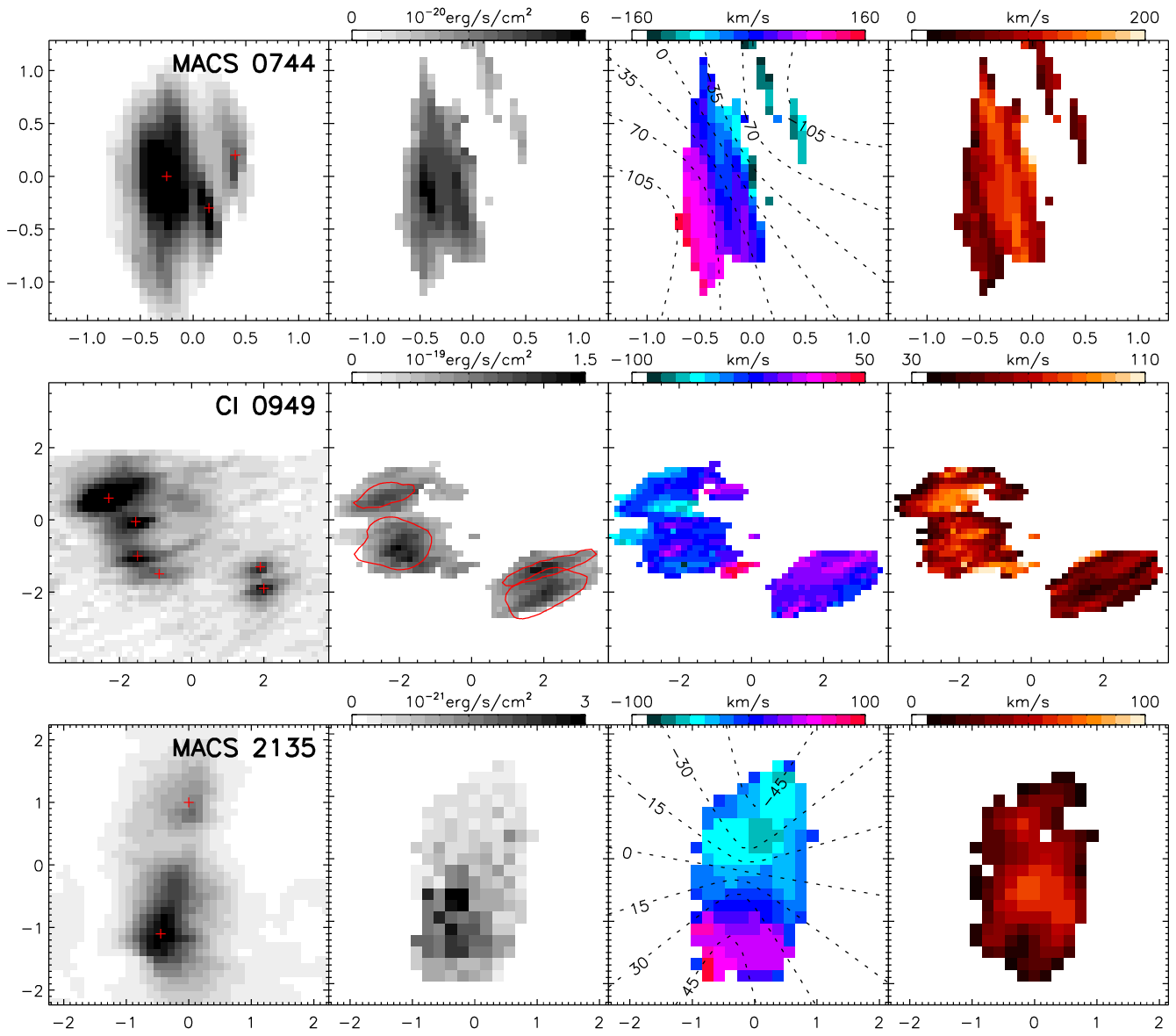,angle=0,width=7in}}
\caption{(continued)}
%\label{fig:osiris_montage}
\end{figure*}

Before we discuss what can be learned from the sample as a whole, we
briefly review the results for each of the six galaxies in turn.

\subsection{Review of Individual Properties}

\subsubsection{Cl\,0024+1709}

After accounting for lensing amplification, the extent of H$\alpha$
emission is $\sim10 \times 18$ kpc from our lens model, making this
the largest object in our sample.  
The integrated H$\alpha$ emission
line flux implies an intrinsic SFR$=27\pm6\Msolyr$ (uncorrected for
dust extinction).

The H$\alpha$ emission line and rest-UV morphologies both show a
clumpy ring surrounding a fainter central region
(Figure~\ref{fig:osiris_montage}), similar to that seen in some other
massive, extended galaxies at high redshift (e.g. the $z=2.26$ galaxy
Q2643-BX482 in \citealt{Genzel08}) as well as local ring galaxies
\citep{Romano08}.  The narrow ring, $\sim$10\,kpc diameter, and
misaligned central feature suggest that the ring of concentrated star
formation was triggered by an on-axis collision with another galaxy,
as opposed to dynamical resonances within the galaxy.
This hypothesis is supported by the numerical simulations
of \cite{Lynds76} and identification of interacting companions to ring
galaxies by \cite{Romano08}.
Unfortunately, the H$\alpha$ emission in the central region coincides
with a detector glitch in most of the IFU exposures and so the central
dispersion and star formation density are only poorly constrained,
with $\sigma=50\pm21 \kms$. In the ring, five UV-
and H$\alpha$-bright knots are resolved with similar isophotal radius,
velocity dispersion and luminosity of $R\simeq0.5$\,kpc (for
$L_{H\alpha}>1.5\times10^{-16}$\,erg\,s$^{-1}$\,cm$^{-2}$\,arcsec$^{-2}$),
$\sigma\simeq 45-65 \kms$, and $L_{H\alpha}\sim2\times10^{-17}$
erg\,s$^{-1}$\,cm$^{-2}$, with uncertainties from the lens model and
emission line fits of $\sim 40$\% in radius, 15\% in dispersion and
30\% in luminosity.  This implies a dynamical mass of the H$\alpha$ clumps from
the velocity dispersion of $M_{dyn}=2\pm1\times10^9\Msol$ assuming
uniform density, and a H$\alpha$ star-formation rate of 
SFR $=2.0\pm0.7\Msolyr$.  The velocity field shows a smooth velocity
gradient around the ring with peak-to-peak amplitude
$2\,V_c\,\sin{i}=120\pm10\,\kms$.  Using the axial ratio from the
optical imaging, we estimate $\cos{i}=0.6\pm0.1$ which suggests a
rotationally supported mass
$M_{dyn}=R\,V_c^2/G=2.1\pm0.8\times10^{10}\,M_{\odot}$
within a radius of 9\,kpc. If the
galaxy has a uniform spherical mass distribution, then the observed
dispersion suggests a larger dynamical mass of $5\times10^{10}\Msol$.
The resolved star forming regions in the ring thus account for about
50\% of the integrated H$\alpha$ flux and 15--25\% of the mass
interior to the ring.

\subsubsection{MACS\,J0451+0006}

The integrated H$\alpha$ emission line flux suggests a star-formation
rate of SFR=$7\pm2 \Msolyr$, with the two brightest clumps
(Figure~\ref{fig:osiris_montage}) contributing $0.8\pm0.2$ and $3.6\pm0.9\,\Msolyr$. 
Some parts of the galaxy were not observed with OSIRIS due to the limited field of view 
and 20 arcsecond extent of the arc. We expect these regions to contribute an 
additional $3\pm1 \Msolyr$ based on their rest-frame UV flux.
The large magnification of this galaxy ($\mu=49\pm11$) leads to
an unprecedented 60\,pc typical resolution in the source plane,
providing unique insight into the kinematic structure of the star forming
regions.
The H$\alpha$ velocity map reveals a coherent velocity field across
the galaxy and within the clumps with a peak to peak amplitude of
$80 \pm20\,\kms$ (Figure~\ref{fig:rotcurves}).
The bi-symmetric velocity field together with the good disk model fit ($\chi_{\nu}^2$=3.6) 
suggest that the kinematics of this system are likely dominated by rotation rather than merging.

\subsubsection{MACS\,J0712+5932}

From our IFU observations, only the brightest UV clump is detected in
H$\alpha$ with an intrinsic star-formation rate of
SFR $=5\pm1 \Msolyr$ and we place limits on the remaining
clumps of SFR $<3 \Msolyr$ ($3\sigma$).  The H$\alpha$ flux in the
brightest clump is consistent with the total galaxy flux derived from
longslit spectroscopy, implying very little star formation in the rest
of the system.  The two prominent UV-bright regions have FWHM sizes
$\lsim0.5$\,kpc and are separated by $1.5$\,kpc
(Figure~\ref{fig:osiris_montage}).  The H$\alpha$ emission in the
brighter region exhibits a velocity gradient of $38\pm8\,\kms$ along
the highly sheared direction.
While the disk model provides an excellent fit to the velocity field
with $\chi_{\nu}^2=0.4$, it is unclear whether the dynamical state of
MACS\,J0712+5932 is dominated by rotation, merging, or dispersion since
only the strongest H$\alpha$-emitting region is detected with OSIRIS.
Assuming a
uniform density within a radius of 0.2\,kpc and $\sigma=80\pm5 \kms$
we derive a dynamical mass of M$_{dyn}=1.4\pm0.2\times10^9 \Msol$
for the nebular emission region.  The N[{\sc ii}]$\lambda6583$
emission line is detected at the $2 \sigma$ level
in individual spatial pixels consistent with the flux ratio
N[{\sc ii}]$\lambda6583$/H$\alpha = 0.2$ observed in NIRSPEC longslit
spectra, with no measurable gradient in the flux ratio.

\subsubsection{MACS\,J0744+3927}

H$\alpha$ emission is detected across $2.0$\,kpc in the source plane
and is resolved into 3 bright clumps in both the rest-frame UV
continuum and H$\alpha$.  The two dimensional velocity map clearly
shows a bi-symmetric velocity field and is well described by the
disk model with $\chi^2_{\nu}=3.8$ and peak to peak amplitude of
$280\pm25$\,km\,s$^{-1}$, the largest observed in any high-redshift
galaxy of this size (Figure~\ref{fig:rotcurves}).
The velocity dispersion varies from $\sim60$--$130\kms$ and peaks near
the dynamical center of the galaxy.
This suggests a turbulent primitive disk with a total star-formation
rate of $2.4\pm0.5 \Msolyr$.
The dynamical mass
within a radius of 1\,kpc is
M$_{dyn}$=$R\,V_c^2/G=5\pm1\times10^9\Msol/\sin^2{i}$ inferred from
the velocity gradient, or M$_{dyn}=11\pm2\times10^9\Msol$ from the
flux-weighted mean dispersion assuming uniform density.

The large rotation velocity, low SFR, and central dispersion peak
conceivably make MACS\,J0744+3927 the most evolved disk-like galaxy of
its size yet observed at high redshift and show that small, low mass
($\sim10^{10}\Msol$), turbulent field disk galaxies are already in
place only 3\,Gyr after the big bang.

\subsubsection{Cl\,0949+5153}

\begin{figure*}
\centerline{\psfig{file=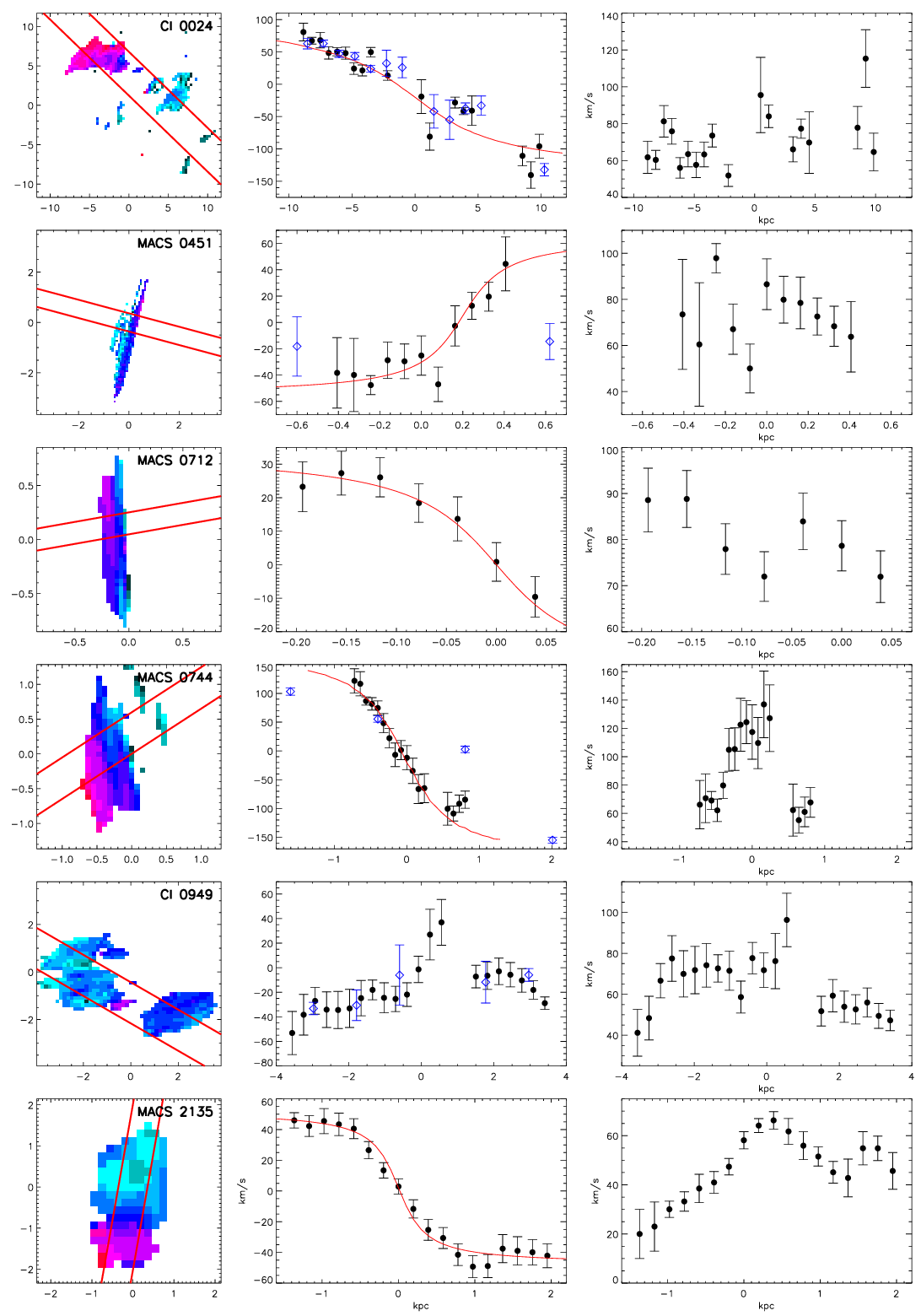,angle=0,height=8.25in}}
\caption{Left to right: velocity field, one dimensional extracted velocity profile, and equivalent velocity dispersion profile. Velocity and dispersion curves are extracted from left to right along the slit shown on the left figures. The rotation curve one would obtain in the absence of lensing with 0.15'' resolution typical of AO-corrected observations is shown by the blue diamonds. Best-fit profiles of the velocity field are shown in red. No model is shown for Cl\,0949+5153 since the velocity field is inconsistent with a rotating disk.}
\label{fig:rotcurves}
\end{figure*}

The reconstructed image of Cl\,0949+5153 at $z=2.393$ shows two
diffuse regions separated by $\simeq3.9$ kpc, with the southwest
component redshifted by $17\pm6\,\kms$ from its companion
(Figure~\ref{fig:rotcurves}).  Both components show resolved
velocity gradients of order 30 and 100\,$\kms$ but in opposite directions,
indicating that these two regions are likely merging (this system can
not be described by a disk model). Each component is further resolved
into multiple clumps of diameter $0.3-0.7$\,kpc within an isophote of
$L_{[OIII]}=1.8\times10^{-16}$erg\,s$^{-1}$\,cm$^{-2}$\,arcsec$^{-2}$.
Interestingly, while the clumpy [O{\sc iii}] morphology appears to
correlate with the rest-frame UV, the ratio of rest-UV to nebular
emission surface brightness varies considerably between clumps.  This
could be caused by non-uniform extinction or different star formation
histories, since the rest-UV traces older star formation episodes than
H$\alpha$ flux ($\lsim100$ and $\lsim20$ Myr respectively).  Assuming
a global H$\beta$/[O{\sc iii}] flux ratio from longslit spectra and
case B recombination, the H$\beta$-derived star formation rate of the
entire system is $20\pm6\,M_{\odot}$\,yr$^{-1}$ with 63\% of the
nebular emission in the larger northeast region.

Dynamically, it appears that Cl\,0949+5153 is the only major merger in
our sample, containing two spatially separated components with
opposite velocity gradients and enhanced dispersion between components.
From the velocity dispersion we crudely estimate that the northeast
and southwest components have mass
$12\times$ and $4\times10^{9}\Msol$ respectively. Both of these
components are fragmented further into multiple large clumps with
seemingly coherent velocity fields.

\subsubsection{MACS\,J2135-0102}

For completeness, we also include in our sample discussion of the
velocity field of LBG\,J2135-0102 (also known as the ``Cosmic Eye''
due to its lensed morphology).  
\cite{Stark08} show that the global dynamics suggest a thick galaxy
disk in an early state of assembly with rotation speed (corrected for
inclination) of $V_{c}=67\pm7\,\kms$ and $V_{c}/\sigma_{r}=1.2\pm0.1$.

The source plane morphology of this galaxy comprises two distinct
star-forming regions embedded in a rotating disk.  The northern
component has a diameter of $\sim0.8$\,kpc, a star formation rate of
$21\pm3\, M_{\odot}$\,yr$^{-1}$ and a peak line width of
$50\pm5$\,km\,s$^{-1}$.  Together, these suggest an intensely star
forming region with dynamical mass
$M_{dyn}=1.2\pm0.5\times10^9\,M_{\odot}$. The southern component is
brighter in the rest-frame UV, but has a lower H$\beta$-inferred star
formation rate of SFR $=11\pm2 \Msolyr$.  The diameter of
$1.3\pm0.3$\,kpc and peak line width of 65\,km\,s$^{-1}$ suggest a
mass $M=3\pm1\times10^9\,M_{\odot}$.

\subsection{Properties of the Ensemble}

Having discussed the sources individually, we now turn to general
conclusions based upon their integrated properties as listed in Table 2.

\begin{table*}
\begin{center}
{\footnotesize
{\centerline{\sc Table 2.}}
{\centerline{\sc Intrinsic Galaxy Properties}}
\begin{tabular}{lccccccccc}
\hline
\noalign{\smallskip}
Galaxy           & Diameter    & $L/L^*$ &     SFR     &       $M_{dyn}$   & $\sigma$ & $V_{max} \sin{i}$ & $i$ & $\chi^2_{\nu}$ \\
                 &  (kpc)      &         & ($\Msolyr$) &   ($10^9 \Msol$)  & ($\kms$) & ($\kms$)   & ($^{\circ}$) & \\
\hline
Cl\,0024+1709    &  20 $\pm$ 2  & 0.7 &  27$\pm$6   & 55$\pm$10            & 69$\pm$5 &  76$\pm$12 & $50^{+10}_{-30}$ & 5.9 \\
MACS\,J0451+0006 &   5 $\pm$ 1  & 0.4 &   7$\pm$2   & 19$\pm$4             & 80$\pm$5 &  38$\pm$10 & $25^{+20}_{-10}$ & 3.6 \\
MACS\,J0712+5932 & 1.5 $\pm$0.3 & 0.2 &   5$\pm$1   & 5.7$\pm$1.2          & 81$\pm$2 &  15$\pm$4  & $40^{+20}_{-25}$ & 0.4 \\
MACS\,J0744+3927 & 2.0 $\pm$0.3 & 0.1 & 2.4$\pm$0.5 & 11$\pm$2             & 99$\pm$4 & 129$\pm$12 & $45^{+15}_{-25}$ & 3.8 \\
Cl\,0949+5152    &   7 $\pm$ 1  & 0.5 &  20$\pm$6   &                      & 66$\pm$3 &   &   & \\
       \,    NE  & 4.0 $\pm$0.6 &     &  13$\pm$5   & 12$\pm$2             & 71$\pm$2 &   &   & \\
       \,    SW  & 2.0 $\pm$0.3 &     &   7$\pm$3   & 3.8$\pm$0.6          & 57$\pm$2 &   &   & \\
MACS\,J2135-0102 & 3.5 $\pm$0.3 & 1.5 &  40$\pm$5   & 5.9$\pm$1.0          & 54$\pm$4 &  47$\pm$7 & $55 \pm 10$ & 3.7 \\
\hline
\label{table:properties}
\end{tabular}
}
\caption{Source-plane properties of the lensed galaxies. Values for
  MACS\,J2135-0102 are reported in Stark et al. (2008).  Diameter is
  the extent of nebular emission except for MACS\,J0712+5932 where it
  represents the extent of detected V-band emission, with uncertainty
  accounting for spatial resolution and errors in the lens model. The
  luminosity is compared to $L^*$ for the LBG population at $z = 3$
  (see text for methodology). Error in luminosity is 15--30\%
  dominated by the uncertain magnification. Velocity dispersion
  $\sigma$ is taken as the flux-weighted mean. $V_{max} \sin{i}$ is
  computed as half the maximum peak-to-peak velocity from fits to the
  velocity field. Inclination $i$ is the value relative to the plane
  of the sky for the best-fit disk model, with uncertainty given as
  the range of parameter space within 1$\sigma$ of the best
  fit. Cl\,0949+5152 is separated into the northeast and southwest
  merging components; no circular velocity is given since rotating
  disk models do not provide a good fit. Dynamical mass is calculated
  as $C\,R\,\sigma^2 / G$ using the diameter and dispersion values
  given in the table with $C = 5$ and error from $R$ and $\sigma$,
  although the true uncertainty in $M_{dyn}$ is roughly a factor
  of two due to unknown mass distribution and rotation support.}
\end{center}
\end{table*}

\subsubsection{Kinematics}
\label{section:kinematics}

All six objects in our sample show well-resolved velocity structure.
A visual inspection of Figures~\ref{fig:osiris_montage} and
\ref{fig:rotcurves} reveals that four galaxies (Cl\,0024+1709,
MACS\,J0451+0006, MACS\,J0744+3927, and MACS\,J2135-0102) have smooth
velocity gradients suggesting ordered rotation whilst Cl\,0949+5153
appears to be a merger with both components showing well-resolved
velocity gradients. The dynamical state of MACS\,J0712+5932 is unclear
since only the brightest H$\alpha$-emitting region is detected with OSIRIS.
The nebular emission shows a smooth velocity gradient in MACS\,J0712+5932,
but subtends only four resolution elements in the most highly
magnified of the three images and so it is unclear wither the
kinematics of MACS\,J0712+5932 are dominated by rotation, dispersion,
or merging.

For the galaxies studied here, the best fit disk models have
$\chi_{\nu}^2$ values which range from 0.4 to 5.9 with a mean of 3.5,
indicating agreement typically within 2$\sigma$ for individual pixels.
For reference, disk models for 18 galaxies in the SINS survey which
show the most prominent rotation yield best-fit $\chi_{\nu}^2$ values
of 0.2--20 to the models of \cite{Cresci09}, so our simple model
provides an equally good fit to the lensed galaxies.  All galaxies
show small-scale deviations from the model as indicated by the typical
$\chi_{\nu}^2 > 1$; these proper motions could be caused by the
effects of gravitational instability, or simply be due to the
unrelaxed dynamical state indicated by high velocity dispersions
$\sigma>50\,\kms$.  We therefore conclude that the model provides an
adequate fit to the data and that the velocity fields are consistent
with the kinematics of a turbulent rotating disk, except in the case
of Cl\,0949+5152.

It is illustrative to demonstrate the dramatic improvement in our understanding of
the internal dynamics of our sources that arises uniquely through the
improved spatial sampling enabled by studying strongly-lensed
galaxies. To do this we bin the source-plane data cubes of our targets to a 
coarser resolution typical of that achievable for {\it unlensed
sources} observed with an AO-corrected resolution of 0.15$''$.
We then re-fit the emission lines and extract one-dimensional velocity
profiles using the same methods as adopted for the original data. Since the
lensing magnification results in $\mu$ times more flux spread over
$\mu$ times as many pixels, the binned signal-to-noise ratio is a
a factor $\sqrt{\mu}$ higher than equivalent observations of
unlensed versions of our galaxies.  These simulations therefore 
represent non-lensed galaxies observed for much longer integrations
($\sim 100$ hours!).  Even so, the resulting velocity profiles 
(also shown in Figure~\ref{fig:rotcurves}) are considerably inferior
to those of our lensed data. A credible rotation curve is only retrieved for
Cl\,0024+1709, our least magnified galaxy, with important kinematic detail
lost in all other objects (for example, MACS\,J0712+5932 is unresolved).
Such poor spatial sampling is insufficient to distinguish between rotation and 
merging. More quantitatively, velocity gradients in all of our objects
except Cl\,0024+1709 are significantly underestimated (typically
$V_{max,lens}/V_{max,non-lens}=0.6\pm0.2$).

\subsubsection{Physical Characteristics: Size, Luminosity and Mass}
\label{section:characteristics}

\begin{figure}
\centerline{\psfig{file=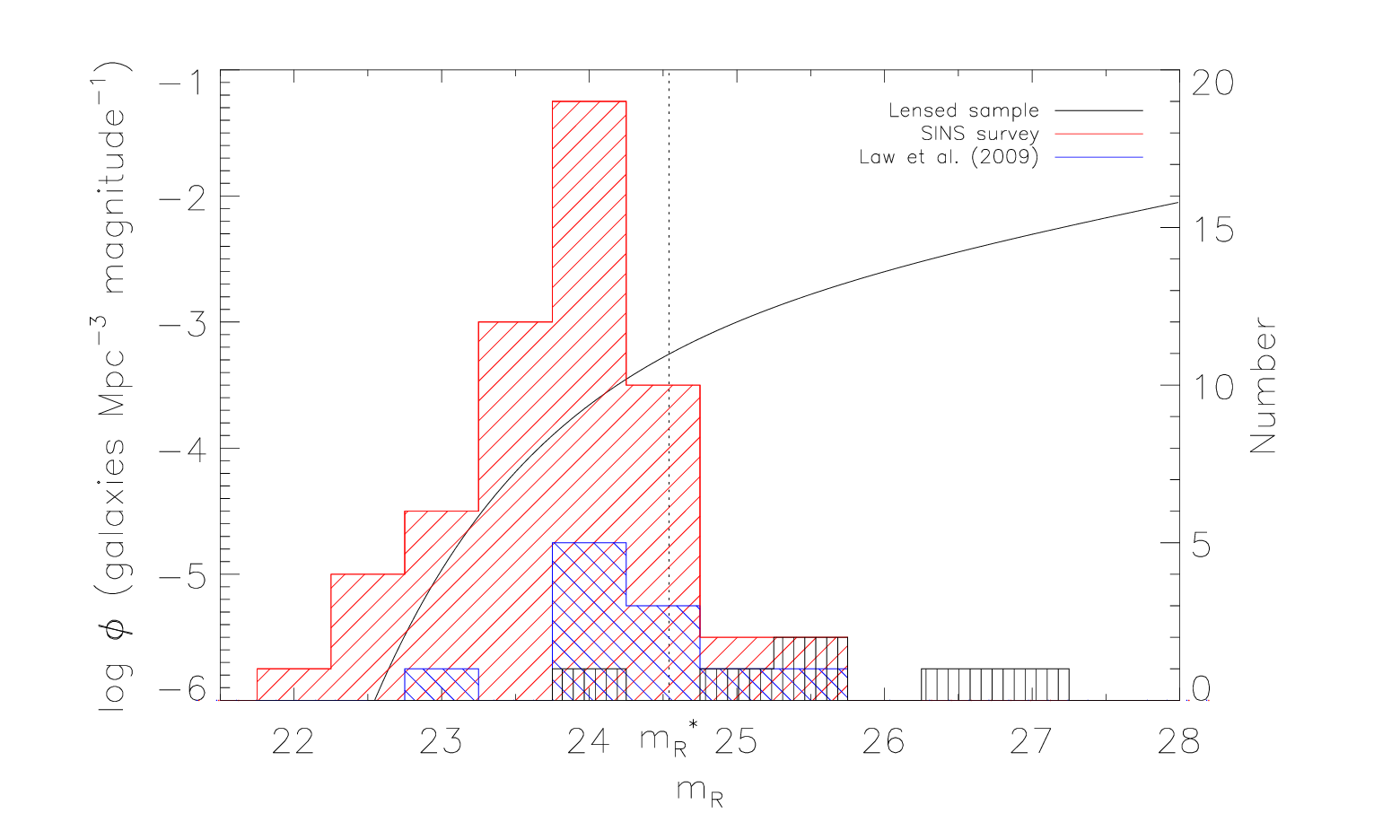,angle=0,width=3.5in}}
\caption{Apparent R-band luminosity function of $z \sim 3$ LBGs from
  Shapley et al. (2001) compared to the distribution of
  suitably-corrected apparent R magnitudes for our lensed sample as
  well as the SINS (Forster Schreiber et al. 2009) and Law et
  al. (2009) samples. We compute the apparent R magnitudes of the IFU samples at $z=3$ for an effective wavelength $\lambda=600$--800 nm depending on the available photometry.}
\label{fig:lf}
\end{figure}

Next, we examine the integrated physical properties of the
lensed galaxies.  First we briefly compare the luminosity, size and
mass to the general Lyman break population at $z\simeq2$--3.  In
Figure~\ref{fig:lf} we compare the distribution of apparent $R$-band
magnitudes for our sample to that of the Lyman-break population at $z
\sim 3$ \citep{Shapley01}.
The comparison demonstrates that five of our six galaxies are fainter than
$L^*$, ranging from $0.1-1.5\,L^*$, with median $0.5\,L^*$
well below that of other spatially resolved IFU studies.
The intrinsic H$\alpha$ flux of the arcs is
also lower than in other surveys, with a mean and median inferred SFR of
17\,$\Msolyr$ (Table 2) compared to 26--33\,$\Msolyr$ in \cite{Law09}
and \cite{Forster09}.
Comparing the
typical spatial extent of the nebular emission to other surveys, the
median radius (extent of detected nebular emission) and dynamical mass
of the lensed sample are 2.1\,kpc and
$1.3\times10^{10}\Msol$, somewhat more extended than the compact galaxies
studied by \cite{Law09} which have median radius of 1.3\,kpc and
M$_{dyn}=0.7\times10^{10}\Msol$.
Turning to the dynamics, the flux-weighted mean velocity dispersions
$\sigma_{mean}$ are perfectly consistent with those studied by Law et
al. (2009): both samples span a range of 50--100 $\kms$ with mean and
median 70--75 $\kms$, indicating that the two sets of galaxies probe
the same range of dynamical mass density.  We note that the median
H$\alpha$ FWHM (from longslit spectra; \citealt{Erb06}) and stellar
mass of $\simeq$6 kpc and $2.9\times10^{10}\Msol$ are a better
representation of the \cite{Law09} sample as they do not depend on
sensitivity of the data.  The SINS survey probes somewhat more
extended objects, with median H$\alpha$ FWHM and
stellar mass of $\simeq$8 kpc and M$_{\star}\simeq3\times10^{10}\Msol$.  
The mass density and extent of nebular emission of the
lensed galaxies are therefore comparable to the more
luminous \cite{Law09} sample and somewhat lower than in the SINS
survey.  The lensed galaxies are also below the average size and mass
for $L^*$ systems at similar redshifts, with the notable exception of
Cl\,0024+1709.

\begin{figure}
\centerline{\psfig{file=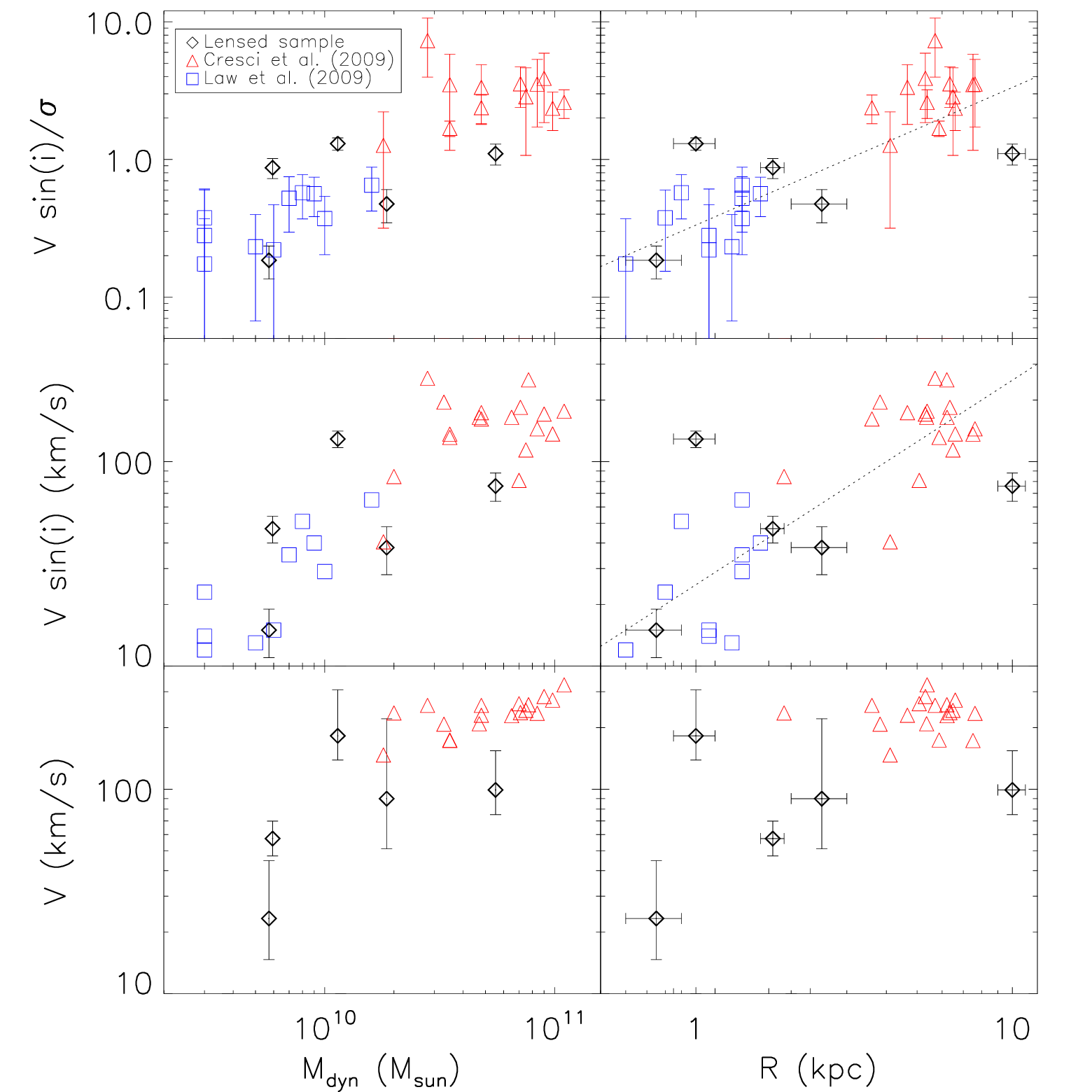,angle=0,width=3.5in}}
\caption{Kinematics as a function of dynamical mass and radius. Values
  for the lensed sample are given in Table~2 and described in the
  text. Data for the comparison samples are taken from Tables~3 and 4
  of Law et al. (2009) and Table~2 of Cresci et al. (2009) and are
  discussed in \S\ref{section:characteristics}. The median $V \sin{i}
  / R = 25 \kms$\,kpc$^{-1}$ is plotted in the middle right panel, and
  the corresponding $V/\sigma$ for typical $\sigma = 75 \kms$ is shown
  in the upper right.}
\label{fig:dynamics}
\end{figure}

Our high resolution data enable us to determine reliable dynamical
properties and compare with other IFU observations. In particular we
can examine the prevalence of dispersion and rotation as a function of
size and dynamical mass. \cite{Law09} report evidence for
dispersion-dominated kinematics in compact low-mass galaxies in
contrast to the $\sim$equal mix of rotation, dispersion, and
complex/merger kinematics found in the SINS survey \citep{Forster09},
although the dispersion-dominated fraction is higher for more compact
SINS galaxies.  However, this claim is hampered by the small number of
resolution elements ($\simeq$2--4) subtended by each source (although see
\citealt{Epinat09}).  We can address this issue with the superior
resolution offered by gravitational lensing.  The relevant relations
between $V/\sigma$ and size or dynamical mass is shown in
Figure~\ref{fig:dynamics} for the five lensed galaxies which are
reasonably well fit by disk models compared to the compact LBGs from
\cite{Law09} and the rotation-dominated SINS galaxies described by
\cite{Cresci09}, showing that the ratio of velocity shear to dispersion is
higher for larger, more massive galaxies.
Furthermore, the general agreement between the lensed sample and other
comparison samples is also shown in Figure~\ref{fig:dynamics} and
demonstrates that the dynamics of sub-$L^*$ star-forming galaxies do
not differ substantially from more luminous objects.

Noting that many galaxies do not reach an asymptotic velocity within
the region probed by our IFU observations, it is likely that larger circular velocities exist in faint outer regions of these galaxies.
Indeed, the spatial extent and mass inferred from longslit spectra and photometry are higher than from relatively shallow IFU data. 
We therefore also compare the observed
velocity gradient and, where appropriate, inclination-corrected
rotational velocity as a function of size and dynamical mass in
Figure~\ref{fig:dynamics}.  The observed velocity gradients of 3/5
lensed galaxies and the entire comparison sample
are within 0.3 dex of the median value $V_{shear}/R=25
\kms$\,kpc$^{-1}$.
The consistent velocity gradients and corresponding $V_{shear}/\sigma$
obtained with a typical $\sigma=75\kms$ (Figure~\ref{fig:dynamics})
suggest that the extended structure in some of
the compact objects may in fact host velocity fields comparable to the
larger disk galaxies.

While MACS\,J0451+0006 and MACS\,J2135-0102 have velocity gradients
consistent with the other samples considered here, the clear outlier
MACS\,J0744+3927 demonstrates that at least some small galaxies at
this redshift have significant dynamical support from ordered
rotation.  This galaxy has an inclination-corrected
$V_c/\sigma=1.8^{+1.2}_{-0.4}$ at radius $R=1$\,kpc and a
H$\alpha$-derived star-formation rate surface density
$\Sigma_{SFR}=0.8\pm0.3$, which is a factor three lower than any
galaxy in the dispersion-dominated sample, possibly indicating a
later stage of evolution.

Motivated by the relatively large velocity shear and small
$\Sigma_{SFR}$ of MACS\,J0744+3927, we explore the possibility that
the low $V/\sigma$ observed in many compact galaxies is an effect of
selecting galaxies in early stages of evolution with large dispersions
and star formation rates. This is expected in models of star formation
feedback, which predict that mechanical energy introduced by supernovae
leads to higher velocity dispersion and hence lower $V/\sigma$ as
$\Sigma_{SFR}$ increases. In one such model, \cite{Dib06} simulate the velocity
dispersion within the inter-stellar medium of galaxies induced by
supernovae energy injection (assuming a 25\% feedback efficiency) and predict a
correlation between star-formation surface density and velocity
dispersion.

\begin{figure}
\centerline{\psfig{file=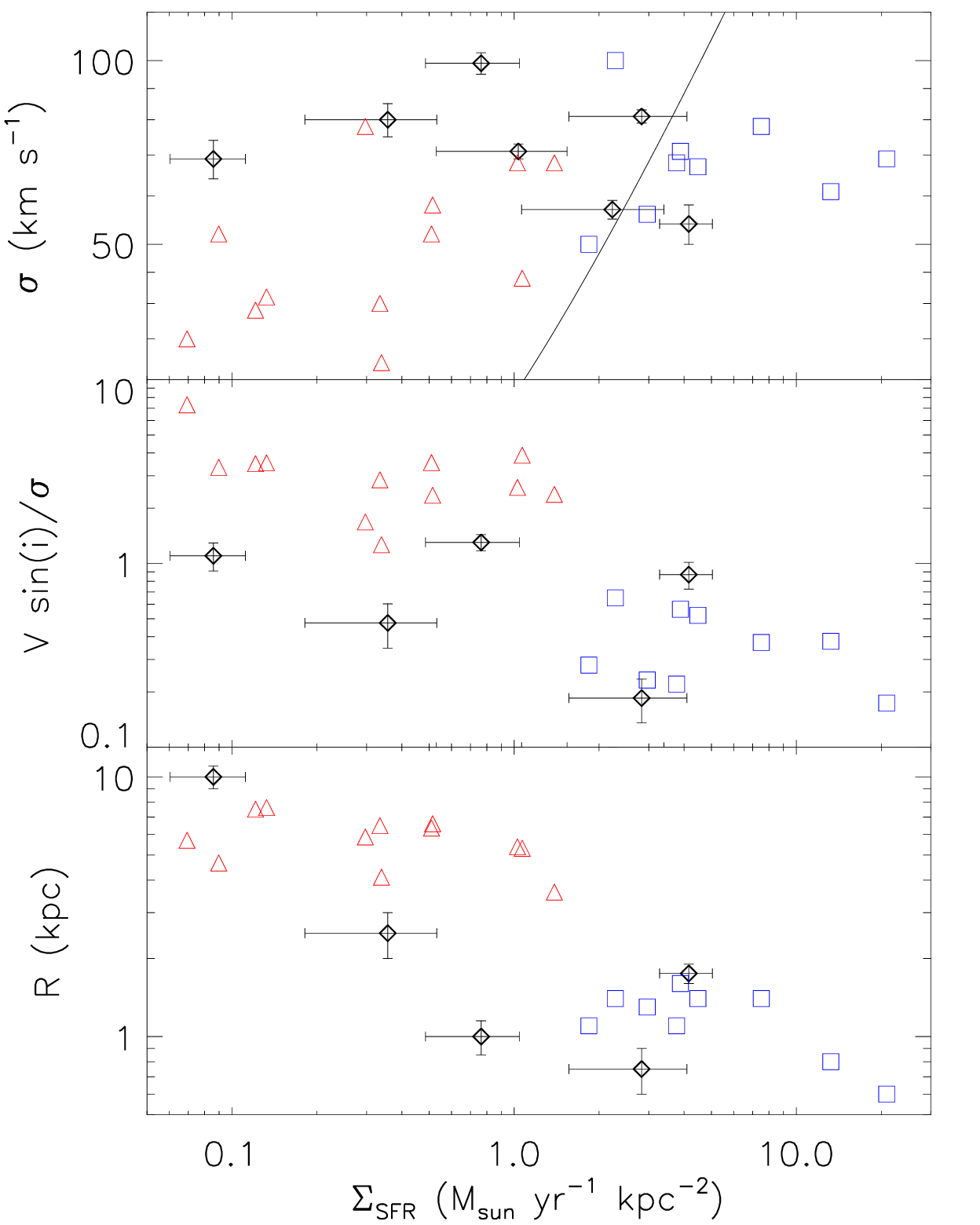,angle=0,width=3.5in}}
\caption{Velocity dispersion, $V_{shear}/\sigma$, and radius as a
  function of star formation rate density estimated from H$\alpha$
  flux. Symbols are the same as in Figure~\ref{fig:dynamics}. The
  solid line in the top panel is a fit to simulations; see text for
  details. The observed $\sigma$ is inconsistent with the
  simulations. Dispersion is at best weakly correlated with
  $\Sigma_{SFR}$. The strong correlation of radius and
  $\sigma/V_{shear}$ with $\Sigma_{SFR}$ is likely due to different
  sensitivity of the observations.}
\label{fig:sfrsigma}
\end{figure}

To test whether supernova feedback might be responsible for the high
dispersion observed in all $z\gsim2$ star-forming galaxies, we show
the velocity dispersion $\sigma$ as a function of $\Sigma_{SFR}$ in
Figure~\ref{fig:sfrsigma}.  For the range of $\Sigma_{SFR}$ with
modeled supernova feedback, the observed velocity dispersions
$\sigma=30$--80$\kms$ are much higher than the predicted
$\sigma=10$--15$\kms$.  We note that the density used in these
simulations is $\sim30\times$ lower than inferred for the $z\simeq2$
galaxies, so the simulated dispersion is likely overestimated by a
factor of $\sim5$.  While the data show a slight trend of $\sigma$
increasing with $\Sigma_{SFR}$, the observed relation is inconsistent
with simulations.  We therefore conclude that supernova feedback is
insufficient to explain the observed velocity dispersions.
Figure~\ref{fig:sfrsigma} shows the relation between
$V/\sigma$--$\Sigma_{SFR}$, clearly demonstrating that $V/\sigma$
decreases with $\Sigma_{SFR}$.  However, this trend is likely
ultimately due to the velocity-size relation: larger galaxies tend to
have larger rotation velocities (Figure~\ref{fig:dynamics}) and
smaller $\Sigma_{SFR}$ (Figure~\ref{fig:sfrsigma}).  This is explained
by different sensitivities of the data, as deeper spectra (lower
$\Sigma_{SFR}$) reveal more extended structures at larger radius with
larger rotation speed.  The velocity-size correlation contributes much
more to the observed $V/\sigma$--$\Sigma_{SFR}$ relation than any
correlation between $\Sigma_{SFR}$ and the velocity dispersion.  These
data thus do not support the hypothesis that $V/\sigma$ is strongly
affected by the density of star formation: $\sigma$ increases by less
than a factor of 2 over two orders of magnitude in $\Sigma_{SFR}$.

\subsection{Star-Formation Scales within Disks}

The small $V_{max}/\sigma \leq 1.8$ and clumpy morphology of all
galaxies suggest that the rotating disks are highly turbulent
and may be dynamically unstable.
We therefore
explore the scale lengths for gravitational collapse within the high
redshift disk galaxies.

Evidence is accumulating that the mode of star formation may be very
different in early systems compared to that seen locally
\citep{Bournaud08,Elmegreen05}.  Rather than forming stars within
giant molecular clouds which condense out of a stable galaxy, star
formation may be triggered by fragmentation of a dynamically unstable
system.  Briefly, in a rotating disk of gas and stars, perturbations
smaller than a critical wavelength $L_{max}$ are stabilized against
the inward pull of gravity by velocity dispersion while those larger
than some $L_{min}$ are stabilized by centrifugal force. If the
dispersion and rotation velocity are too low, $L_{min} > L_{max}$ and
perturbations of intermediate wavelength grow exponentially. This
interplay is summarized by the Toomre parameter $Q=L_{max}/L_{min}$
which is calculated from the velocity dispersion, rotation curve, and
mass distribution \citep{Toomre64}. Galaxies with $Q < 1$ are
therefore unstable on scales between $L_{max}$ and $L_{min}$ and will
fragment into giant dense clumps. This could trigger star formation in
clouds of much higher mass and radius than GMCs in local spiral
galaxies with $Q > 1$, and can explain the clump-cluster and chain
morphologies observed in many high-redshift galaxies.  Dynamical
friction, viscosity and tidal interactions may cause these clumps to
migrate toward the center of the galaxy potential, forming a bulge
which stabilizes the system against further fragmentation.

From the galaxies whose velocity fields can be reasonably well
described by rotating systems, we calculate the Toomre parameter
via:
\begin{equation}
  Q =\frac{\sigma_r\kappa}{\pi G\Sigma}
\end{equation}
which describes the stability of a rotating disk of gas. 
If $Q<1$ the system is unstable to local gravitational collapse and will fragment into overdense clumps.
The value of $\kappa$ is somewhat uncertain as
it depends on the unknown mass distribution; our observations are
consistent with a range $\sqrt{2}\frac{V_c}{R}$--$2\frac{V_c}{R}$
corresponding to constant $V_c$ and $V_c\propto R$
respectively. Adopting $\kappa=\sqrt{3}V_{max}/R$ appropriate for a
uniform disk and using dynamical mass to estimate the surface mass
density $\Sigma$, we find an inclination-corrected $Q\lsim0.6$ for all
galaxies in our sample.  We estimate that the uncertainty in $Q$ is
dominated by a factor of $\simeq2$ error in the dynamical mass.  The
assumed $\kappa$ introduces a negligible 15\% uncertainty, with an
additional random error of $\sim30$\% from the input parameters.
Disk thickness and stellar abundance also affect the value of $Q$.  Combined,
these effects result in roughly a factor of 2 uncertainty.  Even so,
these galaxies all appear to be dynamically unstable since
$Q<1$. Hence we expect them to fragment into massive clumps on scales
of order the Jeans length for dispersion support. In a uniform disk,
the largest scale for which velocity dispersion stabilizes against
gravitational collapse is
\begin{equation}
  L_J=\frac{\pi\sigma^2}{8G\Sigma}
\end{equation}
which is readily estimated from the dispersion and dynamical mass
density. 
As with $Q$, the uncertainty in $L_J$ is a factor of $\simeq2$ dominated by the dynamical mass with additional uncertainty from the unknown mass distribution, disk height, stellar content, and directional dependence of $\sigma$.
The resulting instability scale is
1--3 kpc for Cl\,0024+1709 and 0.1--1.5 kpc for all other objects,
consistent with the observed clump sizes.

\section{Star Forming Regions at High Redshift}

For each galaxy in our sample, the gain in resolution provided by
gravitational lensing reveals multiple resolved giant H{\sc ii}
regions which, as a group, dominate the
integrated star formation rate.  This gain enables us to study, in
more detail than hitherto, the properties of individual star forming
complexes in typical high redshift galaxies. In the following
analysis, we concentrate on those H{\sc ii} regions with extended
bright emission which are sufficiently isolated.  Ideally we would
construct a complete well-defined sample, however, in the exploratory
work presented here we wish to limit the sample to objects for which
the size and flux can be measured robustly. We therefore specifically
exclude H{\sc ii} regions with overlapping emission in the H$\alpha$
emission line maps (such as those in MACS\,J0744+3927).

In brief, H{\sc ii} regions were selected with a flux isophote in the
image plane where the instrumental response is well known.  Thus we
can check whether each H{\sc ii} region is sufficiently resolved by
comparing with the point-spread function produced by a bright tip-tilt
reference star. To measure the area, each H{\sc ii} region was
deconvolved with the point spread function using a method which
accounts for the distorted shapes produced by gravitational
lensing. The intrinsic flux and area were then computed by
reconstructing the H{\sc ii} region in the source plane.

H{\sc ii} regions were selected above a flux isophote corresponding to
$\simeq3\sigma$, where $\sigma$ is the noise level measured in narrow-band
images, constructed by collapsing continuum-subtracted OSIRIS data cubes around the strongest emission feature.  Only
isolated H{\sc ii} regions (in which there is a single local maximum
within the isophote) were selected.  Regions not meeting this
criterion are excluded from this analysis since attempting to
disentangle multiple confused sources would introduce a large
uncertainty. The isophote used varies from 1.4--4.6\,$\times 10^{-16}$
erg\,s$^{-1}$\,cm$^{-2}$\,arcsec$^{-2}$ for different arcs. The
isophote for each selected H{\sc ii} region is shown in
Figure~\ref{fig:osiris_montage}.  Each H{\sc ii} region subtends
0.5--0.9 arcseconds in the image plane, with the notable exception of
the giant clump in MACS\,J0451+0006 which spans 3.2 arcseconds. The
selected H{\sc ii} regions are all well-resolved since the point
spread function has FWHM $\leq 0.2''$ in all observations.

For each H{\sc ii} region we calculated the total nebular line flux
and intrinsic angular size in the image plane. The luminosity of each
H{\sc ii} region was taken as the total flux within the selection
isophote, with the uncertainty determined from the noise level in the
narrow-band image. The H{\sc ii} region size is more complicated to
derive since the effect of the spatial resolution needs to be taken
into consideration. We proceed by making two basic assumptions. First,
we assume the observed flux distribution is a convolution of the
actual distribution with the instrumental point spread function. We
can therefore compute the approximate intrinsic spatial variance by subtracting
the Gaussian PSF variance from the measured variance within the
isophote, $V_{int} = V_{iso} - \sigma^2_{PSF}$. This is equivalent to
subtracting $\sigma =$\,FWHM$/2.35$ in quadrature from Gaussian
profiles, but the procedure is applicable to non-Gaussian asymmetric
distributions. In all cases $V_{iso} > \sigma^2_{PSF}$ confirming that
the H{\sc ii} regions are well resolved: $V_{iso} / \sigma^2_{PSF} =
2.3$ for one H{\sc ii} region and $\geq 3.2$ for all others. Second,
we assume that the isophotal area scales with spatial variance such
that intrinsic surface area $A_{int} = A_{iso}
\frac{V_{int}}{V_{iso}}$, where $A_{iso}$ is defined as the pixel size
multiplied by the number of pixels within the isophote.  Uncertainty
in the isophotal flux ranges from 12--27\% propagated from flux
calibration and noise level in the narrow-band image.  The uncertainty
in $A_{iso}$ is estimated by assuming an error of PSF $\sigma$ for the
isophote diameter. The range of resulting angular diameter ($2
\sqrt{A_{iso}/\pi}$) and relative error is $0.12 \pm 0.06$ to $0.61
\pm 0.08$ arcseconds.

To determine the source plane properties, each H{\sc ii} region is
finally reconstructed using the lens model transformation.  The total
magnification $\mu$ is calculated from the ratio of image- to
source-plane flux. The source plane diameter is then defined as $d = 2
\sqrt{\frac{A_{iso}}{\mu \pi}}$. This method is robust against the
spatially varying magnification of the lensed arcs. Uncertainty in the
lens model has little effect on the diameter but significantly
contributes to the error in flux. From the H$\alpha$ flux (converted
from [O{\sc iii}]$\lambda 5007$ for Cl\,0949+5153) we estimate the
star formation rate using the \cite{Kennicutt98} prescription. The
values range from 0.8--3.6 $\Msolyr$ star formation rate and 0.3--1.0
kpc diameter.

\begin{figure}
\centerline{\psfig{file=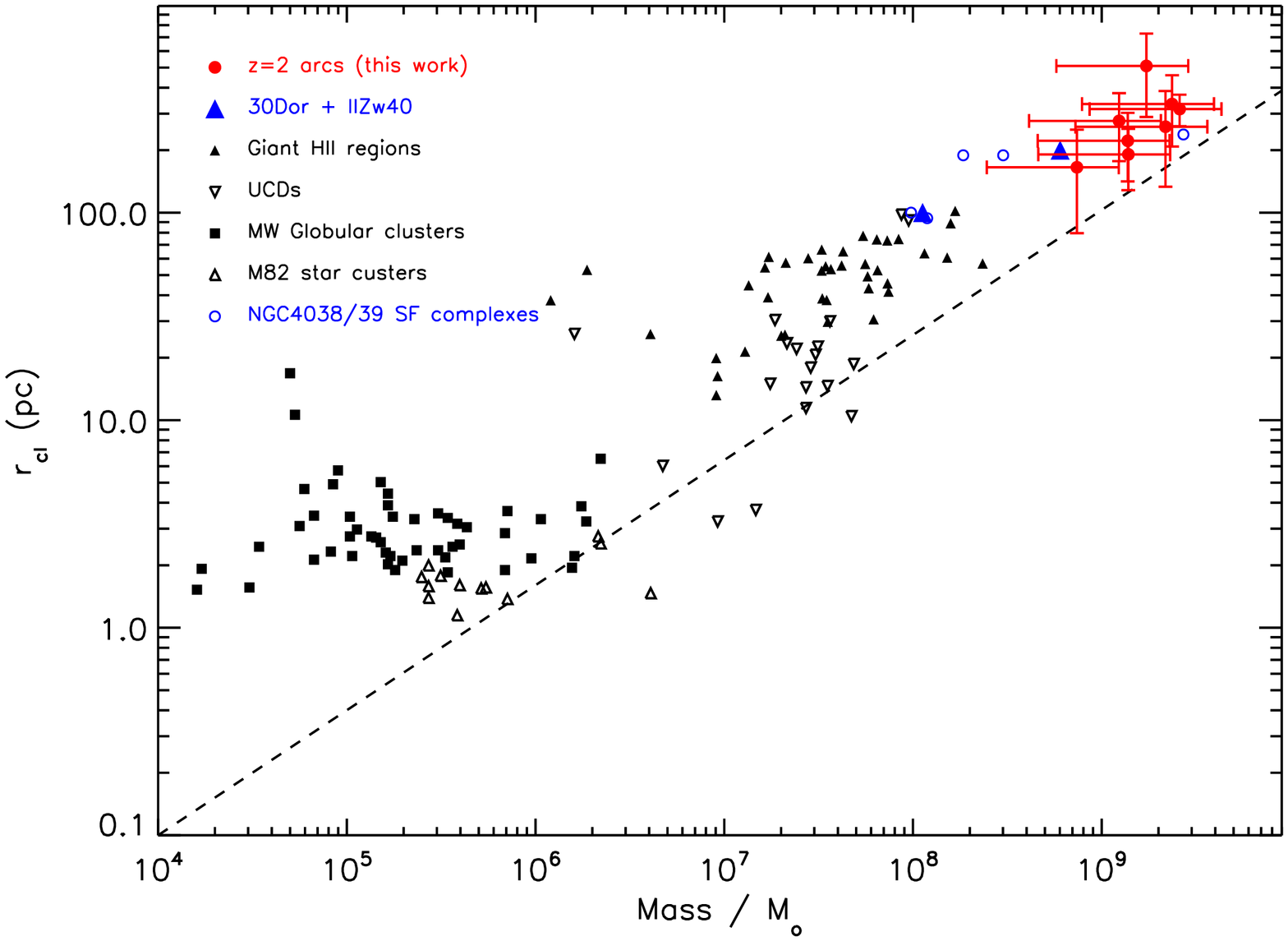,angle=90,width=3.5in}}
\caption{Size versus dynamical mass of individual star-forming regions within our high-redshift sample, compared to local star clusters and H{\sc ii} regions. The high-redshift H{\sc ii} regions have size and mass consistent with the largest star-forming complexes in the local universe. Local data are taken from Harris (1996) and Pryor \& Meylan (1993) for Galactic globular clusters; Ha{\c s}egan et al. (2005), Hilker et al. (2007) and Evstigneeva et al. (2007) for ultra-compact dwarf galaxies; McCrady \& Graham (2007) for M82 clusters; Fuentes-Masip et al. (2000) for giant H{\sc ii} regions; and Bastian et al. (2006) for star-forming complexes in the Antennae.
The dashed line represents a
model for regions which are optically thick to far-infrared
radiation and have undergone adiabatic expansion (Murray 2008). Dynamical mass for the high-redshift data is calculated as $M_{dyn} = C \frac{R\sigma^2}{G}$ with assumed $C = 5$ appropriate for a uniform-density sphere. The mass error bars account for statistical uncertainty in $R$ and $\sigma$.}
\label{fig:clump_mr}
\end{figure}

From the source plane diameters and kinematic data, we estimate the
dynamical mass of the H{\sc ii} regions. We use the flux-weighted mean
velocity dispersion within the clump $\sigma_{mean}$ and assume a
uniform spherical mass distribution ($C=5$) as discussed in \S3.1.
The H{\sc ii}
regions have large flux-weighted mean dispersions, with $\sigma_{mean}$
ranging from 45--80 $\kms$.  The dynamical masses range from
0.7--3\,$\times 10^9 \Msol$, far greater than typical star-forming clusters in
the local universe.  Statistical uncertainty in the dynamical mass is
20--60\%, so the additional factor of $\sim2$ systematic uncertainty in $C$ contributes
significantly to the dynamical mass estimate. The mass-radius relation
of the high-redshift clumps is shown in Figure~\ref{fig:clump_mr} along with a
sample of local star clusters and giant H{\sc ii} regions for comparison. 
This shows that the H{\sc ii} regions studied here are comparable in size and mass to the largest local star-forming complexes and consistent with the mass-radius relation observed locally.

\begin{figure}
\centerline{\psfig{file=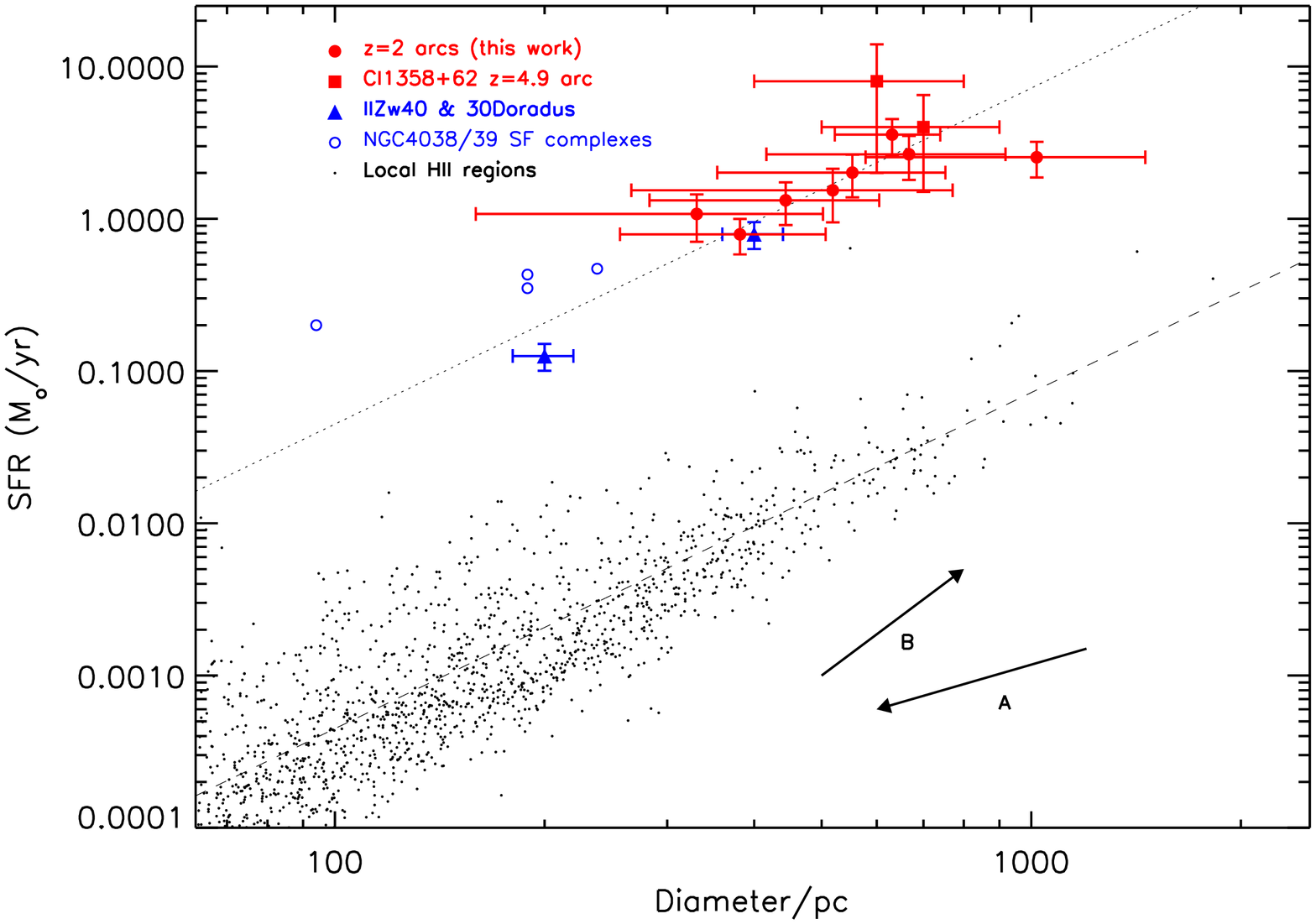,angle=90,width=3.25in}}
\caption{H$\alpha$-inferred star formation rate as a function of H{\sc ii} region size. Red circles represent individual H{\sc ii} regions from our lensed sample. Cl\,0949+5152 was observed in [O{\sc iii}] with the H$\alpha$ flux taken to be 1.2 times the [O{\sc iii}]$\lambda 5007$ flux as determined from long slit NIRSPEC observations. The value for MACS\,J2135-0102 was determined from the H$\beta$ flux assuming case B recombination. 
The star formation rates are all derived from H$\alpha$ luminosities using the Kennicutt (1998) prescription.
Red squares are from the lensed $z=4.92$ galaxy presented by Swinbank et al. (2009), black points are H{\sc ii} regions in local spiral galaxies from Gonzalez-Delgado and Perez (1996), and blue open circles are star cluster complexes in the interacting galaxies NGC 4038/4039 from Bastian et al. (2006). The vectors A and B are explained in the text. The dashed line shows a fit to the data from local spiral galaxies, and the dotted line is the same but increased by a factor of 100$\times$ in SFR density. The high-redshift star-forming regions have significantly higher density of massive stars than H{\sc ii} regions in nearby spirals, but are comparable to the brightest star cluster complexes found in the local universe.}
\label{fig:clump_sfr}
\end{figure}

In Figure~\ref{fig:clump_sfr} we compare the size--luminosity relation
of these star forming regions with equivalent data observed locally
(from \citealt{Gonzalez97} for normal spirals, and \citealt{Bastian06}
for intense starbursts such as the Antennae) and with a lensed arc at $z=4.92$ \citep{Swinbank09}. The sizes observed in
the high redshift galaxies are comparable to the largest local star
forming (H{\sc ii}) regions but with $\sim$100$\times$ higher star
formation rates than in local spiral galaxies.  However, the implied star formation rate
densities are roughly consistent with that observed in the most
vigorous local starbursts.  This result is not significantly affected
by the resolution of different observations, although varying the
selection isophote alters the radius-luminosity relation.
We note that the flux isophotes used to define H{\sc ii} regions in local spirals are much lower, so to quantify this effect we extract individual H{\sc ii} regions from narrow-band H$\alpha$ images of local galaxies from the SINGS and 11HUGS surveys \citep{Kennicutt03,Lee07} using various isophotes.
The effect on the size--luminosity relation as the isophote is increased is shown in Figure~\ref{fig:clump_sfr} as the vector A, which demonstrates that the extracted regions become smaller with higher average surface brightness. The vector B shows the effect of degrading the resolution while the isophote is kept fixed, which causes distinct H{\sc ii} regions to merge together giving the appearance of a larger region with similar surface brightness. Since the surface brightness of the high-redshift H{\sc ii} regions is $\sim 100\times$ higher than in local spirals, we conclude that the offset cannot be explained by different resolution and sensitivity of the various data sets. 
Metallicity also fails to explain the offset, as demonstrated by MACS\,J2135-0102 which has an R$_{23}$ index and C{\sc iv} P Cygni profile suggesting 0.4--1 Z$_{\odot}$ metallicity \citep{Stark08, Quider09}.
This implies that the luminosities in the
high redshift H{\sc ii} regions are truly much larger than in local
spirals. Furthermore
the observed diameters are consistent with the Jeans length for
support by velocity dispersion, suggesting that they collapsed as a
result of disk instabilities. We expect this to occur since the Toomre
Q parameter is less than unity for all galaxies in our sample.
It therefore seems likely that intense star-formation in high redshift galaxies is driven by the fragmentation of gravitationally unstable systems.

\section{Conclusions}

We have utilized strong gravitational lensing together with
laser-assisted guide star adaptive optics at the Keck observatory to
study the internal characteristics of six $z = 1.7-3$ star-forming
galaxies with a spatial resolution down to $\sim 100-200$
parsecs. Extending beyond the diffraction limited capability of
current 8--10 meter class telescopes, our study provides an interesting
preview of the science that will routinely be possible with future 30
meter class optical/infrared telescopes with adaptive optics systems
(Ellis 2009). We demonstrate that enhanced resolution allows us to
resolve morphological and kinematic structure which has not been (and
cannot be) discerned in AO-based surveys of similar non-lensed
sources. In particular we resolve multiple giant star forming regions
of size consistent with that expected from Toomre instability in 
{\sl all} of our targets (including separate merging components).
Furthermore we determine the dynamical state of galaxies which would be too poorly sampled to distinguish between rotation, merging, or dispersion-dominated kinematics without gravitational lensing.

The lensed sample allows a unique study of relatively faint and small galaxies at high resolution. The median luminosity of the sample is half of the characteristic $L^*$ for $z=3$ LBGs and well below that of other high-redshift IFU studies. We find that the kinematics of the sub-$L^*$ population are in general agreement with larger and more luminous galaxies at similar redshifts: four of the six systems are clearly rotating with $V\sin{i}/\sigma = 0.5$--1.3, one has a velocity gradient consistent with rotation within the small H$\alpha$-bright region of the arc detected with OSIRIS, and one is a major merger. All have high velocity dispersion $\sigma = 50$--$100\kms$, consistent with all other resolved observations of $z>2$ star-forming galaxies and comparable to the value $\sigma \simeq 50 \kms$ for stars which formed in the Galactic disk at $z\simeq2$ \citep{Sparke00}. Galaxies with a larger radius of detected nebular emission tend to have higher velocity shear and $V/\sigma$ and lower $\Sigma_{SFR}$; these trends are consistent with other observations.
The case of MACS\,J0744+3927 with inclination-corrected $V=180^{+100}_{-40} \kms$ clearly demonstrates that some small ($R=1.0\pm0.2$\,kpc), fast-rotating field disk galaxies are already in place by $z=2$. Well-sampled velocity fields demonstrate that the ubiquitous giant clumps are often embedded in common rotating systems and are not independent merging components. In contrast, non-lensed clumpy systems with velocity shear have been interpreted as mergers or remained ambiguous. These results demonstrate the need of high-resolution observations to distinguish rotating systems and mergers, as well as to probe the $< 1$\,kpc scale of star formation at high redshift.

The observations presented here represent the best current probe of the scale of star formation at $z \gtrsim 2$ and, as such, can be compared to theoretical predictions. Most of the star formation in our sample occurs within giant H{\sc ii} regions of diameter 0.3--1 kpc comparable to the largest local H{\sc ii} regions, with star formation rate density comparable to the most vigorous local starbursts and $\sim 100 \times$ higher than in typical spiral galaxies. This offset in $\Sigma_{SFR}$ cannot be explained by the different resolution or sensitivity of low- and high-redshift observations. Only one of our six objects is a clear major merger with both merging components broken into high-SFR clumps, suggesting that star formation episodes in sub-L$^*$ high redshift galaxies are triggered by Toomre instability independent of major merger events. We note that the merging system demonstrates no enhancement in star formation (as traced by nebular emission) compared to the non-merging galaxies.  The increased high-mass star formation relative to local spirals likely results from some combination of higher gas density, increased star formation efficiency, shorter star formation timescales, and possibly other effects. The clumpy star formation, high $\Sigma_{SFR}$, kinematics consistent with rotation, and small $Q<1$ are in qualitative agreement with the galaxy evolution model of \cite{Dekel09} in which galaxies accrete their baryonic mass from cold streams. In this model the clumps form due to Toomre instability in turbulent disks and migrate into the galaxy center on timescales of a few hundred Myr, forming a bulge which stabilizes the system and increases the star formation timescale after $\sim 2$ Gyr.
Numerical simulations suggest that cold-stream accretion is a dominant mechanism in galaxy assembly, especially at early times and in galaxies with halo masses less than a few $10^{11} \Msol$ \citep{Keres09,Ocvirk08,Brooks09}. In particular the high-resolution simulations of \cite{Brooks09} show that accretion of cold gas dominates the buildup of stellar disks and bulges in sub-$L^*$ galaxies and allows disk formation at earlier times than if the accreted gas is shock-heated.
The rotating kinematics and distribution of star formation in the lensed sample therefore supports, at least qualitatively, a cold-stream accretion scenario for galaxy formation.

Analysis of our sample has shown the benefit of using lensed sources
to probe the inner structure of early star-forming galaxies. The
advent of many panoramic multi-color imaging surveys will hopefully
reveal many further examples of such systems, ensuring further
progress with laser-assisted guide star adaptive optics. Ultimately,
the next generation of 30-meter optical/near-infrared telescopes
will extend the high-resolution studies presented herein to the
entire population of high-redshift star forming galaxies.

\section*{Acknowledgments}

We thank Randy Campbell, Mark Kassis, Jim Lyke, and Hien Tran for their dedication and assistance in obtaining these observations.
AMS gratefully acknowledges a Royal Astronomical Society Sir Norman
Lockyer Research Fellowships and a Royal Society travel grant. RSE
acknowledges financial support from the Royal Society. JR
acknowledges support from an EU Marie Curie fellowship. DPS
acknowledges support from an STFC Postdoctoral Research Fellowship.
These observations were obtained at the W.M. Keck Observatory, which is
operated as a scientific partnership among the California Institute of
Technology, the University of California and the National Aeronautics
and Space Administration. The Observatory was made possible by the
generous financial support of the W.M. Keck Foundation.

\bibliographystyle{apj}
\bibliography{ref}

\begin{thebibliography}{19}
\expandafter\ifx\csname natexlab\endcsname\relax\def\natexlab#1{#1}\fi

\bibitem[Bastian et al.(2006)]{Bastian06} Bastian, N., Emsellem, E., Kissler-Patig, M., \& Maraston, C.\ 2006, \aap, 445, 471 

\bibitem[Bouch{\'e} et al.(2007)]{Bouche07} Bouch{\'e}, N., et 
al.\ 2007, \apj, 671, 303 

\bibitem[Bournaud et al.(2008)]{Bournaud08} Bournaud, F., et al.\ 2008, \aap, 486, 741 

\bibitem[Broadhurst et al.(2000)]{Broadhurst00} Broadhurst, T., Huang, X., Frye, B., \& Ellis, R.\ 2000, \apjl, 534, L15 

\bibitem[Brooks et al.(2009)]{Brooks09} Brooks, A.~M., Governato, F., Quinn, T., Brook, C.~B., \& Wadsley, J.\ 2009, \apj, 694, 396 

\bibitem[Colley et al.(1996)]{Colley96} Colley, W.~N., Tyson, J.~A., \& Turner, E.~L.\ 1996, \apjl, 461, L83 

\bibitem[Coppin et al.(2007)]{Coppin07} Coppin, K.~E.~K., et al.\ 2007, \apj, 665, 936 

\bibitem[Courteau(1997)]{Courteau97} Courteau, S.\ 1997, \aj, 114, 2402

\bibitem[Cresci et al.(2009)]{Cresci09} Cresci, G., et al.\ 2009, \apj, 697, 115 

\bibitem[Dekel et al.(2009)]{Dekel09} Dekel, A., Sari, R., \& Ceverino, D.\ 2009, \apj, 703, 785 

\bibitem[Dib et al.(2006)]{Dib06} Dib, S., Bell, E., \& Burkert, A.\ 2006, \apj, 638, 797 

\bibitem[Dickinson et al.(2003)]{Dickinson03} Dickinson, M., Papovich, C., Ferguson, H.~C., \& Budav{\'a}ri, T.\ 2003, \apj, 587, 25 

\bibitem[Dye et al.(2007)]{Dye07} Dye, S., Smail, I., Swinbank, A.~M., Ebeling, H., \& Edge, A.~C.\ 2007, \mnras, 379, 308 

\bibitem[Ebeling et al.(2001)]{Ebeling01} Ebeling, H., Edge, A.~C., \& Henry, J.~P.\ 2001, \apj, 553, 668 

\bibitem[El{\'{\i}}asd{\'o}ttir et al.(2007)]{Eliasdottir07} El{\'{\i}}asd{\'o}ttir, {\'A}., et al.\ 2007, arXiv:0710.5636 

\bibitem[Elmegreen \& Elmegreen(2005)]{Elmegreen05} Elmegreen, B.~G., \& Elmegreen, D.~M.\ 2005, \apj, 627, 632 

\bibitem[Elmegreen et al.(2009)]{Elmegreen09} Elmegreen, B.~G., Elmegreen, D.~M., Fernandez, M.~X., \& Lemonias, J.~J.\ 2009, \apj, 692, 12 

\bibitem[Epinat et al.(2009)]{Epinat09} Epinat, B., Amram, P., Balkowski, C., \& Marcelin, M.\ 2009, arXiv:0904.3891 

\bibitem[Erb et al.(2006)]{Erb06} Erb, D.~K., Steidel, C.~C., Shapley, A.~E., Pettini, M., Reddy, N.~A., \& Adelberger, K.~L.\ 2006, \apj, 646, 107 

\bibitem[Evstigneeva et al.(2007)]{2007AJ....133.1722E} Evstigneeva, E.~A., Gregg, M.~D., Drinkwater, M.~J., \& Hilker, M.\ 2007, \aj, 133, 1722 

\bibitem[F{\"o}rster Schreiber et al.(2006)]{Forster06} F{\"o}rster Schreiber, N.~M., et al.\ 2006, \apj, 645, 1062 

\bibitem[F{\"o}rster Schreiber et al.(2009)]{Forster09} F{\"o}rster Schreiber, N.~M., et al.\ 2009, arXiv:0903.1872 

\bibitem[Fuentes-Masip et al.(2000)]{2000AJ....120..752F} Fuentes-Masip, O., Mu{\~n}oz-Tu{\~n}{\'o}n, C., Casta{\~n}eda, H.~O., \& Tenorio-Tagle, G.\ 2000, \aj, 120, 752 

\bibitem[Genzel et al.(2006)]{Genzel06} Genzel, R., et al.\ 2006, \nat, 442, 786 

\bibitem[Genzel et al.(2008)]{Genzel08} Genzel, R., et al.\ 2008, \apj, 687, 59 

\bibitem[Gonzalez Delgado \& Perez(1997)]{Gonzalez97} Gonzalez Delgado, R.~M., \& Perez, E.\ 1997, \apjs, 108, 199 

\bibitem[Harris(1996)]{1996yCat.7195....0H} Harris, W.~E.\ 1996, VizieR Online Data Catalog, 7195, 0 

\bibitem[Ha{\c s}egan et al.(2005)]{Hasegan05} Ha{\c s}egan, M., et al.\ 2005, \apj, 627, 203 

\bibitem[Hilker et al.(2007)]{2007A&A...463..119H} Hilker, M., Baumgardt, H., Infante, L., Drinkwater, M., Evstigneeva, E., \& Gregg, M.\ 2007, \aap, 463, 119 

\bibitem[Hopkins \& Beacom(2006)]{Hopkins06} Hopkins, A.~M., \& Beacom, J.~F.\ 2006, \apj, 651, 142 

\bibitem[Jullo et al.(2007)]{Jullo07} Jullo, E., Kneib, J.-P., Limousin, M., El{\'{\i}}asd{\'o}ttir, {\'A}., Marshall, P.~J., \& Verdugo, T.\ 2007, New Journal of Physics, 9, 447 

\bibitem[Kennicutt(1998)]{Kennicutt98} Kennicutt, R.~C., Jr.\ 1998, \apj, 498, 541 

\bibitem[Kennicutt et al.(2003)]{Kennicutt03} Kennicutt, R.~C., Jr., et al.\ 2003, \pasp, 115, 928 

\bibitem[Kere{\v s} et al.(2009)]{Keres09} Kere{\v s}, D., Katz, N., Fardal, M., Dav{\'e}, R., \& Weinberg, D.~H.\ 2009, \mnras, 395, 160 

\bibitem[Kneib et al.(1993)]{Kneib93} Kneib, J.~P., Mellier, Y., Fort, B., \& Mathez, G.\ 1993, \aap, 273, 367 

\bibitem[Kneib et al.(1996)]{Kneib96} Kneib, J.-P., Ellis, R.~S., Smail, I., Couch, W.~J., \& Sharples, R.~M.\ 1996, \apj, 471, 643 

\bibitem[Larkin et al.(2006)]{Larkin06} Larkin, J., et al.\ 2006, New Astronomy Review, 50, 362 

\bibitem[Law et al.(2007a)]{Law07a} Law, D.~R., Steidel, C.~C., Erb, D.~K., Pettini, M., Reddy, N.~A., Shapley, A.~E., Adelberger, K.~L., \& Simenc, D.~J.\ 2007, \apj, 656, 1 

\bibitem[Law et al.(2007b)]{Law07} Law, D.~R., Steidel, C.~C., Erb, D.~K., Larkin, J.~E., Pettini, M., Shapley, A.~E., \& Wright, S.~A.\ 2007, \apj, 669, 929 

\bibitem[Law et al.(2009)]{Law09} Law, D.~R., Steidel, C.~C., Erb, D.~K., Larkin, J.~E., Pettini, M., Shapley, A.~E., \& Wright, S.~A.\ 2009, \apj, 697, 2057 

\bibitem[Lee et al.(2007)]{Lee07} Lee, J.~C., Kennicutt, R.~C., Funes, S.~J., Jos{\'e} G., Sakai, S., \& Akiyama, S.\ 2007, \apjl, 671, L113 

\bibitem[Lehnert et al.(2009)]{Lehnert09} Lehnert, M.~D., Nesvadba, N.~P.~H., Tiran, L.~L., Matteo, P.~D., van Driel, W., Douglas, L.~S., Chemin, L., \& Bournaud, F.\ 2009, \apj, 699, 1660 

\bibitem[Limousin et al.(2009)]{Limousin09} Limousin, M., et al.\ 2009, in preparation

\bibitem[Lynds \& Toomre(1976)]{Lynds76} Lynds, R., \& Toomre, A.\ 1976, \apj, 209, 382 

\bibitem[McCrady \& Graham(2007)]{McCrady07} McCrady, N., \& Graham, J.~R.\ 2007, \apj, 663, 844 

\bibitem[Murray(2009)]{Murray09} Murray, N.\ 2009, \apj, 691, 946 

\bibitem[Nesvadba et al.(2006)]{Nesvadba06} Nesvadba, N.~P.~H., et al.\ 2006, \apj, 650, 661 

\bibitem[Ocvirk et al.(2008)]{Ocvirk08} Ocvirk, P., Pichon, C., \& Teyssier, R.\ 2008, \mnras, 390, 1326 

\bibitem[Pryor \& Meylan(1993)]{Pryor93} Pryor, C., \& Meylan, G.\ 1993, Structure and Dynamics of Globular Clusters, 50, 357 

\bibitem[Quider et al.(2009)]{Quider09} Quider, A.~M., Shapley, A.~E., Pettini, M., Steidel, C.~C., \& Stark, D.~P.\ 2009, arXiv:0910.0840 

\bibitem[Reddy \& Steidel(2004)]{Reddy04} Reddy, N.~A., \& Steidel, C.~C.\ 2004, \apjl, 603, L13 

\bibitem[Richard et al.(2007)]{Richard07} Richard, J., et al.\ 2007, \apj, 662, 781 

\bibitem[Richard et al.(2009)]{Richard09} Richard, J., et al.\ 2009, in preparation

\bibitem[Romano et al.(2008)]{Romano08} Romano, R., Mayya, Y.~D., \& Vorobyov, E.~I.\ 2008, \aj, 136, 1259 

\bibitem[Sand et al.(2005)]{Sand05} Sand, D.~J., Treu, T., Ellis, R.~S., \& Smith, G.~P.\ 2005, \apj, 627, 32 

\bibitem[Santos et al.(2004)]{Santos04} Santos, M.~R., Ellis, R.~S., Kneib, J.-P., Richard, J., \& Kuijken, K.\ 2004, \apj, 606, 683 

\bibitem[Shapley et al.(2001)]{Shapley01} Shapley, A.~E., Steidel, C.~C., Adelberger, K.~L., Dickinson, M., Giavalisco, M., \& Pettini, M.\ 2001, \apj, 562, 95 

\bibitem[Shapley et al.(2003)]{Shapley03} Shapley, A.~E., Steidel, C.~C., Pettini, M., \& Adelberger, K.~L.\ 2003, \apj, 588, 65

\bibitem[Siana et al.(2009)]{Siana09} Siana, B., et al.\ 2009, \apj, 698, 1273 

\bibitem[Smith et al.(2005)]{Smith05} Smith, G.~P., Kneib, J.-P., Smail, I., Mazzotta, P., Ebeling, H., \& Czoske, O.\ 2005, \mnras, 359, 417 

\bibitem[Sparke \& Gallagher(2000)]{Sparke00} Sparke, L.~S., \& Gallagher, J.~S., III 2000, Galaxies in the Universe, by Linda S.~Sparke and John S.~Gallagher, III, pp.~416.~ISBN 0521592410.~Cambridge, UK: Cambridge University Press, September 2000.,  

\bibitem[Spergel et al.(2003)]{Spergel03} Spergel, D.~N., et al.\ 2003, \apjs, 148, 175 

\bibitem[Stark et al.(2007)]{Stark07} Stark, D.~P., Ellis, R.~S., Richard, J., Kneib, J.-P., Smith, G.~P., \& Santos, M.~R.\ 2007, \apj, 663, 10 

\bibitem[Stark et al.(2008)]{Stark08} Stark, D.~P., Swinbank, A.~M., Ellis, R.~S., Dye, S., Smail, I.~R., \& Richard, J.\ 2008, \nat, 455, 775 

\bibitem[Swinbank et al.(2007)]{Swinbank07} Swinbank, A.~M., Bower, R.~G., Smith, G.~P., Wilman, R.~J., Smail, I., Ellis, R.~S., Morris, S.~L., \& Kneib, J.-P.\ 2007, \mnras, 376, 479 

\bibitem[Swinbank et al.(2009)]{Swinbank09} Swinbank, M., et al.\ 2009, arXiv:0909.0111 

\bibitem[Tacconi et al.(2008)]{Tacconi08} Tacconi, L.~J., et al.\ 
2008, \apj, 680, 246 

\bibitem[Toomre(1964)]{Toomre64} Toomre, A.\ 1964, \apj, 139, 1217 

\bibitem[Veilleux \& Osterbrock(1987)]{Veilleux87} Veilleux, S., \& Osterbrock, D.~E.\ 1987, \apjs, 63, 295 

\bibitem[Wizinowich et al.(2006)]{Wizinowich06} Wizinowich, P.~L., et al.\ 2006, \pasp, 118, 297 

\bibitem[Wright et al.(2009)]{Wright08} Wright, S.~A., Larkin, J.~E., Law, D.~R., Steidel, C.~C., Shapley, A.~E., \& Erb, D.~K.\ 2009, \apj, 699, 421 



\end{thebibliography}
\bsp

\end{document}